\documentclass[10pt]{article}
\usepackage[letterpaper, margin=1in]{geometry} 
\usepackage{authblk}
\usepackage{url}
\usepackage{listings} 
\usepackage{graphicx} 
\usepackage{amsmath}  
\usepackage{amsfonts} 
\usepackage{mathrsfs}
\usepackage{mathtools}
\usepackage{multicol} 
\usepackage{multirow} 
\usepackage{makecell} 
\usepackage{booktabs}
\usepackage{float} 
\usepackage{tikz} 
\usepackage{hyperref} 
\usepackage{nameref}
\usepackage{lipsum}
\usepackage{comment}
\usepackage{caption}
\usepackage{cite} 
\usepackage{subcaption} 
\usepackage [english]{babel}
\usepackage [autostyle, english = american]{csquotes}
\MakeOuterQuote{"}
\usepackage[normalem]{ulem}
\usepackage{xcolor}

\usepackage[labelfont=bf,labelsep=period,justification=raggedright,singlelinecheck=off]{caption}
\usepackage[figurename=Fig]{caption}

\makeatletter
\renewcommand{\@biblabel}[1]{\quad#1.}
\makeatother

\title{Gene communities in co-expression networks across different tissues}

\author[1]{Madison Russell}
\author[2]{Alber Aqil}
\author[3]{Marie Saitou}
\author[2]{Omer Gokcumen}
\author[1,4]{Naoki Masuda}

\affil[1]{Department of  Mathematics, University at Buffalo}
\affil[2]{Department of Biological Sciences, University at Buffalo}
\affil[3]{Faculty of Biosciences, Norwegian University of Life Sciences}
\affil[4]{Institute for Artificial Intelligence and Data Science, University at Buffalo}

\begin{document}
\date{}
\maketitle

\begin{abstract}
With the recent availability of tissue-specific gene expression data, e.g., provided by the GTEx Consortium, there is interest in comparing gene co-expression patterns across tissues. One promising approach to this problem is to use a multilayer network analysis framework and perform multilayer community detection. Communities in gene co-expression networks reveal groups of genes similarly expressed across individuals, potentially involved in related biological processes responding to specific environmental stimuli or sharing common regulatory variations. We construct a multilayer network in which each of the four layers is an exocrine gland tissue-specific gene co-expression network. We develop methods for multilayer community detection with correlation matrix input and an appropriate null model. Our correlation matrix input method identifies five groups of genes that are similarly co-expressed in multiple tissues (a community that spans multiple layers, which we call a generalist community) and two groups of genes that are co-expressed in just one tissue (a community that lies primarily within just one layer, which we call a specialist community). We further found gene co-expression communities where the genes physically cluster across the genome significantly more than expected by chance (on chromosomes 1 and 11). This clustering hints at underlying regulatory elements determining similar expression patterns across individuals and cell types. We suggest that \textit{KRTAP3-1}, \textit{KRTAP3-3}, and \textit{KRTAP3-5} share regulatory elements in skin and pancreas. Furthermore, we find that \textit{CELA3A} and \textit{CELA3B} share associated expression quantitative trait loci in the pancreas. The results indicate that our multilayer community detection method for correlation matrix input extracts biologically interesting communities of genes.
\end{abstract}


\section*{Author Summary}
Genes that are similarly expressed across individuals (i.e., co-expressed) are potentially involved in related biological processes. Therefore, the identification and biological analysis of co-expressed genes may be useful for revealing genes associated with specific diseases or other phenotypes. Because gene co-expression depends on the tissue in general, we compared co-expression patterns across four different exocrine gland tissues. This problem lends itself to multilayer network analysis in which each layer of the multilayer network is a tissue-specific gene co-expression network. The nodes in the network represent genes, and a pair of genes is directly connected by an edge if the two genes are co-expressed. We developed a method to detect groups of co-expressed genes in the multilayer gene co-expression network using correlational tissue-specific gene expression data. We found some groups of genes that are co-expressed in all four tissues and other groups of genes that are only co-expressed in one tissue. We also found that some of these groups of genes contain genes that are physically clustered across the genome. Our methods reveal groups of genes with potentially different mechanisms of gene co-expression.

\section{Introduction\label{sec:introduction}}
In networks, communities, or modules, are broadly defined as groups of nodes with higher internal than external density of edges compared to a null model \cite{newman2004detecting, fortunato2010community}. 
There have been proposed numerous objective functions to be optimized and algorithms for community detection in networks. Because edges in networks represent a relationship between the nodes, it follows that these communities are groups of nodes that likely share common properties or play a similar role within the network.
Many real-world networks naturally divide into communities, including biological networks, and studying communities
is expected to help us better understand complex biological interactions \cite{hartwell1999molecular, snel2002identification, spirin2003protein, barabasi2004network, barabasi2011network, loscalzo2017network}.

Communities in gene networks are often called gene modules \cite{snel2002identification, spirin2003protein, barabasi2004network}. Methods to find functional gene modules are useful tools for discovering how the genes interact and coordinate to perform specific biological functions \cite{butte1999mutual, bar2003computational, saelens2018comprehensive, kakati2019comparison}. 
Furthermore, studying the relationships between gene modules may reveal a higher-order organization of the transcriptome \cite{langfelder2007eigengene, oldham2008functional}. Biological analyses of gene modules can suggest genes that play a regulatory role in disease or other phenotypes, or identify novel therapeutic target genes for future intervention studies \cite{gargalovic2006identification, van2018gene, gerring2019gene, li2019identification}. 
Additionally, one can study gene modules across evolutionary time to find biologically important groups of co-regulated genes because genes that must be co-expressed together will be under evolutionary pressure to maintain their coordinated expression \cite{wong2005learning, ovens2021comparative}.

While there are various definitions of gene modules, or communities, in gene co-expression networks, gene modules are sets of genes that are similarly expressed across individuals and, therefore, potentially involved in related biological processes \cite{stuart2003gene, wong2005learning, van2018gene}. In such networks, the nodes represent genes, and a pair of nodes is directly connected with each other by an undirected edge if the two genes are co-expressed, i.e., if they show a similar expression pattern across samples \cite{butte1999mutual, stuart2003gene, zhang2005general, gargalovic2006identification, chowdhury2019differential}. Biologically, co-expressed genes may occur because transcription factors may have unique DNA binding sites located in promoter regions of distinct sets of genes \cite{allocco2004quantifying, ribeiro2021molecular},
polymerase binding may cause synchronous transcription of several genes \cite{ebisuya2008ripples}, physically closeby genes may cluster within similarly regulated topologically associated domains \cite{hurst2004evolutionary, sproul2005role, perry2022snake}, or particular environmental factors may concurrently affect genes in a particular pathway~\cite{holter2000fundamental, carter2004gene, barah2016transcriptional}, among other reasons \cite{gaiteri2014beyond}. 
%
Non-biological effects such as batch processing and RNA quality also contribute to gene co-expression \cite{leek2007capturing, leek2010tackling}. In general, one cannot distinguish between the biological and non-biological sources of co-expression from the expression data alone; thus, interpreting co-expression networks is challenging \cite{gaiteri2014beyond, nguyen2019maninetcluster}. 
However, gene co-expression network analysis may be able to clarify novel molecular mechanisms that are relevant to disease and facilitate identification of potential targets for intervention studies \cite{gaiteri2014beyond, van2018gene}. Crucially, gene co-expression and gene expression carry different information. For example, differential co-expression analysis identified the alpha synuclein variant (aSynL) in several Parkinson's disease data sets. In contrast, differential expression analysis alone did not identify this variant since aSynL was highly differentially co-expressed but not highly differentially expressed \cite{rhinn2012alternative}. Gene co-expression analyses can provide novel insights that are likely overlooked or undetected in traditional gene expression analyses \cite{gaiteri2014beyond}.

Gene expression and co-expression may depend on regulatory elements in the genome, which are often specific to different cell types \cite{piro2011atlas, pierson2015sharing, gerring2019gene, saitou2020functional, azevedo2021multilayer}. 
The increased availability of tissue-specific gene expression data allows us to compare and contrast gene expression and co-expression and their communities across different tissues. A challenge for deciphering such data is integrating and distinguishing between communities found in various cell types, determining their biological relevance, and identifying regulatory elements maintaining these communities. For example, a simultaneous analysis of both generic multi-tissue co-expression (derived from aggregated gene expression data from multiple tissues) and tissue-specific co-expression resulted in a more efficient prediction of human disease genes than the use of generic multi-tissue co-expression alone \cite{piro2011atlas}. It has also been found that modules conserved across different types of tissues are likely to have functions common to those tissues \cite{pierson2015sharing, ritchie2016scalable}. In contrast, modules upregulated in a particular tissue are often involved in tissue-specific functions \cite{pierson2015sharing}.

\begin{figure}[t]
\centering
\includegraphics[width=\linewidth]{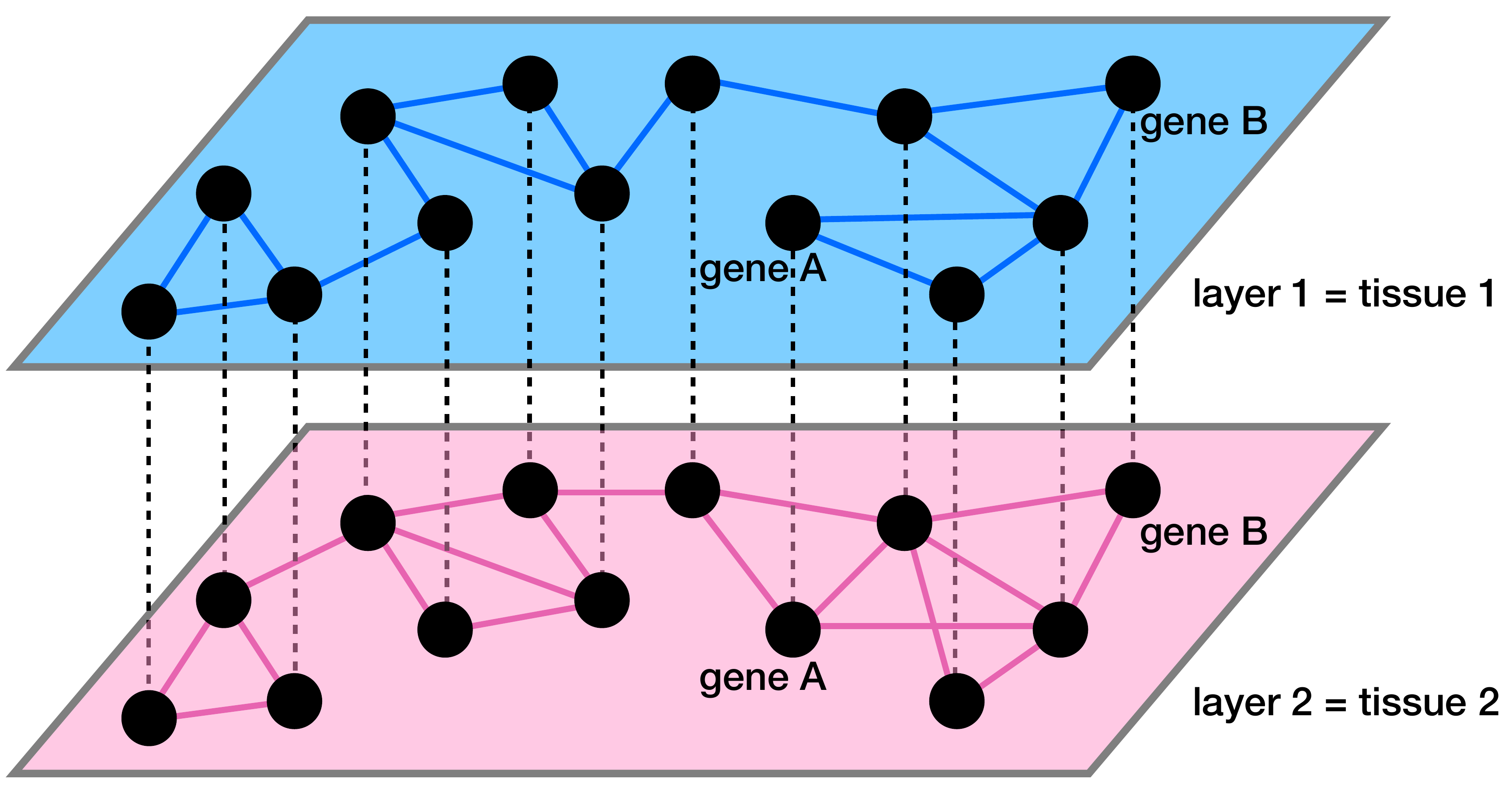}
\caption{{\bf Schematic of a multilayer gene co-expression network.}
  The intralayer edges, shown by the solid lines, represent co-expression. The interlayer edges, shown by the dashed lines, connect the same gene across layers.}
\label{fig:schematic}
\end{figure}

One can regard a set of co-expression networks of genes constructed for multiple tissues as a multilayer network. As we schematically show in Fig~\ref{fig:schematic}, each layer of the multilayer network is a tissue-specific gene co-expression network. The edges within a layer (i.e., intralayer edges) represent tissue-specific co-expression. The edges between the layers (i.e., interlayer edges) connect the same gene across tissues.
Multilayer network analysis, particularly multilayer community detection \cite{mucha2010community, magnani2021community},
is becoming an increasingly popular tool in biological data analysis given that biological systems are often multi-dimensional and involve complex interactions \cite{li2011integrative, haas2017designing, gosak2018network, hammoud2020multilayer}. 
Analyzing single-layer networks separately may be insufficient to reveal the patterns of these complex biological interactions \cite{gosak2018network}. 
For example, multilayer gene co-expression networks, in which each layer consists of a subset of gene pairs with a similar co-expression level, were constructed for comparing healthy and breast cancer co-expression patterns \cite{dorantes2020multilayer}. 
In the healthy multilayer co-expression network, the layers gradually attain hub nodes as one goes towards the top layer, whereas in the breast cancer multilayer network, the majority of layers contain no hub nodes and only a few top layers abruptly start to contain hub nodes \cite{dorantes2020multilayer}. In another application to breast cancer data, a multilayer gene co-expression network in which each layer corresponds to a clinical stage of breast cancer was analyzed \cite{ma2023layer}. A community detection algorithm designed to identify layer-specific modules in multilayer networks finds gene modules in the breast cancer network significantly associated with the survival time of patients \cite{ma2023layer}. Community detection in multilayer stochastic block models, in which each layer is a gene co-expression network at a specific developmental time, reveals different biological processes active at different stages of a monkey's brain development \cite{lei2022bias, zhang2023stochastic}. A Higher-Order Generalized Singular Value Decomposition method allows for simultaneous identification of both ``common'' and ``differential'' modules across several tissue-specific gene co-expression networks \cite{xiao2014multi}. A study of the relationships between communities across different tissue-specific layers of a multilayer gene co-expression network provides promise for our better understanding of inter-tissue regulatory mechanisms through both intra-tissue and inter-tissue transcriptome analysis \cite{azevedo2021multilayer}.

Another application for multilayer approaches is to categorize diseases and drug targets. For instance, analyses of densely connected subgraphs that consistently appear in different layers have revealed disease modules (i.e., groups of diseases extracted from a four-layer disease similarity network in which a node is a disease and the four layers are constructed from protein-protein interaction (PPI), a symptom data set, Gene Ontology, and Disease Ontology) \cite{yu2019conserved} and drug-target modules (i.e., groups of genes extracted from a multilayer network in which each layer is a tissue-specific PPI network) \cite{yu2020exploring}. Groups of diseases that have molecular and phenotypic similarities were discovered in an analysis of a bilayer network of human diseases consisting of a genotype-based and phenotype-based layers~\cite{halu2019multiplex}. 
A multilayer network analysis in which each layer is a similarity matrix among 26 different populations for a given structural variant revealed evolutionarily adaptive structural variants \cite{saitou2022similarity}. Regulatory and signaling mechanisms associated with a given cellular response were discovered using a multilayer community detection method designed for identifying active modules in weighted gene co-expression networks~\cite{li2021active}.
Community detection on tissue-specific multilayer networks composed of a co-expression network, transcription factor co-targeting network, microRNA co-targeting network, and PPI network revealed candidate driver cancer genes \cite{cantini2015detection}. 

As discussed above, the study of co-expression networks can lead to various biological insights \cite{zhang2005general, horvath2008geometric, gaiteri2014beyond}. 
However, there are some limitations to this approach.
Edges of co-expression networks are correlational in nature. In general, creating unweighted or weighted networks from correlation data can be straightforward (e.g., thresholding on the edge weight and/or assuming no edges between negatively correlated node pairs). However, such straightforward methods are subject to various problems such as false positives \cite{barzel2013network, fiecas2013quantifying}, arbitrariness in setting the parameter value such as the threshold on the edge weight \cite{rubinov2011weight, garrison2015stability},
and loss of information by subthreshold or negative correlation values \cite{rubinov2011weight, de2017topological}. Existing methods to estimate sparse networks from correlation matrix data, such as graphical lasso \cite{meinshausen2006high, yuan2007model, friedman2008sparse} or estimation of sparse covariance matrices \cite{schafer2005shrinkage, bien2011sparse, kojaku2019constructing}, mitigate some of these problems.
In contrast to constructing sparse networks, in the present study, we explore the adaptation of network analysis methods to directly work on correlation matrix input.
Such methods have been developed for community detection via modularity maximization \cite{macmahon2015community, bazzi2016community, masuda2018configuration} and clustering coefficients \cite{masuda2018clustering}. A key observation exploited in these studies is that one needs to use appropriate null models for correlation matrices, which are different from those for general networks. In particular, the standard null model for general networks called the configuration model is not a correlation or covariance matrix in general~\cite{macmahon2015community}. In this study, we expand this line of approach to the case of multilayer correlation matrix data.
In particular, we develop a method for community detection by combining multilayer modularity maximization and a configuration model of correlation matrices.
We also develop statistical methods to calculate the significance of each detected community.
We apply our methods to multilayer Pearson correlation matrices representing co-expression of genes in four tissues to compare communities of genes across different tissues.
Code for running our multilayer community detection method with covariance matrix input is available at Github \cite{githubrepo}.

\section{Methods} \label{methods}
\subsection{Data} \label{thedata}
The Genotype-Tissue Expression (GTEx) portal provides open-access tissue-specific gene expression data \cite{lonsdale2013genotype}. For the analyses in the present work, we use the gene transcripts per million (TPM) data from release V8 for four exocrine glands: pancreas, minor salivary gland, mammary gland, and skin (not sun exposed). 
In this pilot study, we limit our analysis to four tissues. We chose these tissues because they are all tissues that interact with the outside world and may have adaptively evolved to different environmental conditions. Specifically, the pancreas plays a vital role in the digestive system, secreting digestive enzymes \cite{brannon1990adaptation}. The salivary gland is the main gatekeeper of our body and contributes to the oral proteome \cite{thamadilok2020human}. The mammary gland produces milk containing immunologic agents to nourish and protect young offspring \cite{mcclellan2008evolution}. The skin protects the body against pathogens, regulates body temperature, and has changed most drastically in human lineage \cite{quillen2019shades, starr2021evolutionary}. Consequently, we hypothesized that these tissues would retain a high level of variation in gene expression levels.

There are $328$ samples from the pancreas, $162$ samples from the minor salivary gland, $459$ samples from the mammary gland, and $604$ samples from the skin (not sun exposed) in this TPM data. Each sample contains gene expression data for $56,200$ different genes. 

The number of genes is much larger than the number of samples for all tissues. Therefore, we focused on a subset of genes for our analysis around the same size as the number of samples in our data, as in \cite{pierson2015sharing,lyu2018condition}. To subset the genes, we identified the top $75$ genes with the highest variance of TPM across all samples \cite{zhang2005general}, separately for each tissue. We chose the number $75$ because the union of the top $75$ genes in terms of the variance of TPM across the four tissues contains $203$ genes, which is not much larger than the smallest number of samples (162 samples). It is well known that estimation of correlation matrices from data is unreliable if the number of elements (i.e., genes in the present case) is comparable with or larger than the number of samples \cite{bun2017cleaning}. 
Nevertheless, to further validate our choice for the number of genes, we repeated some analysis on an expanded network with $371$ genes. We found that the expanded network produces a similar type of partition as the original network, supporting the robustness of our analysis with respect to the number of genes selected for our analysis (see Text A in \nameref{S1Text}).

We looked at the most variable genes because, again, our goal is to understand the underlying genetic and environmental bases of gene expression variation. In fact, most of the highly variably expressed genes are also highly expressed genes. To show this, for each tissue, we calculate the Jaccard index between the top $75$ genes in terms of average TPM and the top $75$ genes in terms of variance of TPM. The Jaccard index is defined as the size of the intersection of two finite sets $A$ and $B$ divided by the size of the union of $A$ and $B$ \cite{jaccard1912distribution}. The range of the Jaccard index is $0$ to $1$, and a larger Jaccard index implies a greater overlap between the two sets of genes. 
We also examine the average rank of the top $75$ genes in variance among all $56,200$ genes. We compute the rank in terms of the average TPM. Therefore, if the average rank is high (i.e., a low number), then the highly variable genes are also relatively highly expressed. We show in
Table~\ref{table:highvar_highavg} the Jaccard index and the average rank of the top $75$ genes for each tissue.
The table indicates that the Jaccard index is at least $0.402$ and the average rank is at most $167.8$.
These results suggest that the top $75$ genes in terms of variance of TPM are overall highly expressed genes as well because we have calculated these indices for $75$ genes in comparison to the $56,200$ genes. This finding is consistent with an established understanding that sequence read count data follows a negative binomial distribution \cite{robinson2007moderated, anders2010differential, conesa2016survey}.

 \begin{table}[!ht]
\centering
\caption{
{\bf Similarity between the highly variable genes and the highly expressed genes in each tissue.}}
\begin{tabular}[b]{|c | c c|}\hline
Tissue & \thead{Jaccard \\ index} & \thead{Average \\ rank} \\\hline
pancreas & $0.685$ & $52.81$ \\
salivary gland & $0.531$ & $64.31$ \\
mammary gland & $0.402$ & $133.5$ \\
skin & $0.442$ & $167.8$ \\\hline
 \end{tabular}
\begin{flushleft} We calculate the Jaccard index between the top $75$ genes in terms of average TPM and the top $75$ genes in terms of variance of TPM. We calculate the average rank of the top $75$ genes in variance, where the rank is in terms of average TPM.
\end{flushleft}
 \label{table:highvar_highavg}
\end{table}

We analyze four-layer networks composed of the $203$ genes in the union of the top $75$ genes in terms of the variance of TPM across the four tissues. We note that the number of nodes must be the same in each layer for our multilayer community detection method described in section~\ref{commdetectioncorr}.

\subsection{Multilayer network construction} \label{multinetworkconstruction}
For each of the four tissues, we generate a $203\times 203$ gene co-expression matrix in which the $(i,j)$-th entry is the Pearson correlation coefficient between the log-normalized TPM of gene $i$ and the log-normalized TPM of gene $j$ across all samples from that tissue. We take the logarithm of TPM before calculating the Pearson correlation coefficient to suppress the effect of outliers; TPM is extremely large for some samples. Let $S$ denote the number of samples from tissue $\alpha$. We denote by $x_{i, \alpha, s}$ and $x_{j, \alpha, s}$ the TPM value for gene $i$ and $j$, respectively, for sample $s\in\{1,2,\ldots,S\}$ in tissue $\alpha$. Then, we calculated the Pearson correlation coefficient between $\log(x_{i, \alpha, s}+1)$ and $\log(x_{j, \alpha, s}+1)$ across the $S$ samples as the co-expression between gene $i$ and gene $j$ in tissue $\alpha$. In other words, we calculate
\begin{equation} \label{eq:genecorr}
r_{\alpha}(i,j)=\frac{\sum_{s=1}^{S}[\log(x_{i, \alpha, s}+1)-m_{i, \alpha}][\log(x_{j, \alpha, s}+1)-m_{j, \alpha}]} {\sqrt{\sum_{s=1}^{S}[\log(x_{i, \alpha, s}+1)-m_{i, \alpha}]^{2}\sum_{s=1}^{S}[\log(x_{j, \alpha, s}+1)-m_{j, \alpha}]^{2}}},
\end{equation}
where 
\begin{equation}\label{eq:gene_m1}
m_{i, \alpha}=\frac{1}{S}\sum_{s=1}^{S}\log(x_{i, \alpha, s}+1)
\end{equation}
and
\begin{equation}\label{eq:gene_m2}
m_{j, \alpha}=\frac{1}{S}\sum_{s=1}^{S}\log(x_{j, \alpha, s}+1).
\end{equation}
We took the logarithm of $x_{i, \alpha, s}+1$ because, in this manner, $x_{i, \alpha, s}=0$ is mapped to $0$.
 
To compare the gene co-expression patterns across the different tissues, we view the four correlation matrices as a four-layer correlation matrix, or categorical layers of a multilayer gene co-expression network. Because the set of genes is the same in the four layers, we place an interlayer edge between the same gene in each pair of layers (i.e., tissues) as shown by the dashed lines in Fig~\ref{fig:schematic}. Therefore, our network is a multiplex network with diagonal and categorical interlayer couplings, where, by definition, the interlayer edges connect each gene with itself in each other layer \cite{kivela2014multilayer, bianconi2018multilayer}.  

We denote the strength of the interlayer coupling that connects node $i$ in layer $\alpha$ to node $i$ in layer $\beta$ as $\omega_{i\alpha\beta}$ \cite{mucha2010community}. One typically assumes that $\omega_{i\alpha\beta}$ takes binary values $\{0,\omega\}$, where $\omega$ is a parameter indicating the absence (i.e., $0$) or presence (i.e., $\omega$) of interlayer edges \cite{mucha2010community}. However, how to set and interpret the $\omega$ value is 
not straightforward \cite{hanteer2020unspoken}. In this work, we use the empirical co-expression (i.e., Pearson correlation coefficient) of gene $i$ between tissues $\alpha$ and $\beta$ as $\omega_{i\alpha\beta}$. Specifically, $\omega_{i\alpha\beta}$ is equal to the right-hand side of Eq.~\eqref{eq:genecorr} with $x_{j, \alpha, s}$ and $m_{j, \alpha}$ being replaced by $x_{i, \beta, s}$ and $m_{i, \beta}$, respectively, and with $S$ being interpreted as the number of samples common to tissues $\alpha$ and $\beta$. Since the majority of studies on multilayer modularity maximization assume non-negative interlayer edge weights, if the obtained $\omega_{i\alpha\beta}$ is negative, we force $\omega_{i\alpha\beta} = 0$. However, note that some studies do include negative interlayer edge weights \cite{zhang2017modularity}.
%

\subsection{Community detection in conventional multilayer networks} \label{communitydetection}
We are interested in detecting communities (also called modules and gene sets) in our multilayer networks to find sets of genes that are similarly expressed across individuals and therefore potentially involved in related biological processes. Some algorithms can detect communities that span between multiple layers as well as communities that lie within just one layer. We are interested in these different types of communities and their biological implications.
A common method to find such communities in multilayer networks is to maximize an objective function called the multilayer modularity \cite{mucha2010community}. However, our multilayer gene networks are based on correlation. Therefore, we develop multilayer modularity for multilayer correlation matrices.
In this section, we review multilayer modularity for usual multilayer networks as a primer to the multilayer modularity for correlation matrices.


The modularity for single-layer undirected networks, which may be weighted, is given by \cite{newman2004finding,newman2004analysis}
\begin{equation} \label{eq:genmod}
Q=\frac{1}{2M}\sum_{i=1}^{N}\sum_{j=1}^{N} \left(A_{ij}- \gamma \frac{k_i k_j}{2M}\right)\delta(g_{i},g_{j}), 
\end{equation}
where $N$ is the number of nodes in the given network; 
$A_{ij}$ is the $(i, j)$-th entry of the adjacency matrix and we assume $A_{ii}=0$ $\forall i \in \{1, \ldots, N\}$; $M=\frac{1}{2}\sum_{i=1}^{N}\sum_{j=1}^{N}A_{ij}$ is the number of edges in the case of unweighted networks and the total weight of all edges in the case of weighted networks;
$\gamma$ is the resolution parameter controlling the size of typical communities found by modularity maximization \cite{fortunato2007resolution}; a large $\gamma$ tends to lead to relatively many small communities;
$k_i k_j/2M$ is equal to the probability that an edge exists, or alternatively the expected edge weight, between nodes $i$ and $j$ under the configuration model; $k_i=\sum_{j=1}^{N}A_{ij}$ is the (weighted) degree of node $i$; $g_{i}$ is the community to which node $i$ belongs; $\delta(g_{i},g_{j})=1$ if $g_{i}=g_{j}$ and $\delta(g_{i},g_{j})=0$ otherwise.

To generalize the modularity to the case of multilayer networks, let $\mathcal{L}$ be the number of layers in the multilayer network. We let $A_{ij\alpha}$ be the $(i, j)$-th entry of the intralayer adjacency matrix, which may be weighted, in network layer $\alpha$. We assume $A_{ii\alpha} = 0$ $\forall i \in \{1, \ldots, N\}$ and $\forall \alpha\in \{1, \ldots, \mathcal{L}\}$. We remind that $\omega_{i\alpha\beta}$ is the weight of the interlayer coupling between node $i$ in layer $\alpha$ and node $i$ itself in layer $\beta$.
The multilayer modularity is given by \cite{mucha2010community}
\begin{equation} \label{eq:multimod}
Q=\frac{1}{2\mu}\sum_{i=1}^{N}\sum_{j=1}^{N}\sum_{\alpha=1}^{\mathcal{L}}\sum_{\beta=1}^{\mathcal{L}}\left[\underbrace{\left(A_{ij\alpha}-\gamma_{\alpha}\frac{k_{i\alpha}k_{j\alpha}}{2m_{\alpha}}\right)\delta_{\alpha\beta}}_{\substack{\text{intralayer}}}+\underbrace{\omega_{i\alpha\beta}\delta_{ij}}_{\substack{\text{interlayer}}} \right]\delta(g_{i\alpha},g_{j\beta}),
\end{equation}
where $k_{i\alpha}=\sum_{j=1}^{N}A_{ij\alpha}$ is the strength (i.e., weighted degree) of node $i$ in layer $\alpha$, and $m_{\alpha}=\frac{1}{2}\sum_{i=1}^{N}k_{i\alpha}$ is the total edge weight in layer $\alpha$. We set $2\mu=\sum_{i=1}^{N}\sum_{\alpha=1}^{\mathcal{L}}(k_{i\alpha} + \sum_{\beta=1}^{\mathcal{L}}\omega_{i\alpha\beta})$, which is equal to twice of the total edge weight. Let $\gamma_{\alpha}$ be the resolution parameter in layer $\alpha$; $\delta_{\alpha\beta}=1$ if $\alpha=\beta$ and $\delta_{\alpha\beta}=0$ otherwise; $\delta_{ij}$ is defined in the same manner; and $g_{i\alpha}$ is the community to which node $i$ in layer $\alpha$ belongs. 
Eq.~\eqref{eq:multimod} implies that communities that contain interlayer edges are rewarded with higher modularity values.

We will discuss the selection of $\gamma_{\alpha}$ in section~\ref{champ}. We use the Louvain algorithm for multilayer modularity maximization. Specifically, we use the iterated GenLouvain function from GenLouvain version 2.2, which repeatedly implements GenLouvain until convergence to an output partition (i.e., until the output partition does not change between two successive iterations) \cite{blondel2008fast, genlouvain}. 

The modularity function $Q$ typically has many local maxima \cite{good2010performance}. Reflecting this fact, most modularity maximization algorithms are stochastic and do not output a unique answer. A common approach to combine the results from multiple partitions of nodes is consensus clustering to obtain a consensus partition \cite{strehl2002cluster}. 
We use the consensus clustering algorithm described in \cite{bassett2013robust} and implemented in the Python package netneurotools version 0.2.3 \cite{netneurotools-package}.
%

\subsection{Community detection in multilayer correlation matrices} \label{commdetectioncorr}
In this section, we expand modularity maximization for correlation matrices
\cite{macmahon2015community, bazzi2016community} to the case of multilayer correlation matrices.

Let $\rho = (\rho_{ij})$ be an $N\times N$ correlation matrix and $\langle\rho\rangle$ be a null model of the correlation matrix of the same size. The modularity for a single correlation matrix is given by
\begin{equation} \label{eq:corrmod}
Q=\frac{1}{C_{\text{norm}}}\sum_{i=1}^{N} \sum_{j=1}^{N} (\rho_{ij}-\langle\rho_{ij}\rangle)\delta(g_{i},g_{j}),
\end{equation}
where $C_{\text{norm}}=\sum_{i=1}^{N} \sum_{j=1}^{N} \rho_{ij}$ is a normalization constant. One can use a modularity maximization algorithm to maximize $Q$ given $\langle\rho\rangle$.

We generalize Eq.~\eqref{eq:corrmod} to the case of a multilayer correlation matrix by writing down an equation in the same form as Eq.~\eqref{eq:multimod}. We will use the term node to refer to a gene in a specific layer of the four-layer correlation matrix. Let $\rho_{ij\alpha}$ be the empirical Pearson correlation coefficient between nodes $i$ and $j$ in layer $\alpha$, and let $\langle\rho_{ij\alpha}\rangle$ be the correlation between nodes $i$ and $j$ in layer $\alpha$ in the null model of the correlation matrix. 
Then, the modularity of a multilayer correlation matrix is 
\begin{equation} \label{eq:corrmultimod}
Q=\frac{1}{C_{\text{norm}}}\sum_{i=1}^{N}\sum_{j=1}^{N}\sum_{\alpha=1}^{\mathcal{L}}\sum_{\beta=1}^{\mathcal{L}}\left[(\rho_{ij\alpha}-\gamma_{\alpha}\langle\rho_{ij\alpha}\rangle)\delta_{\alpha\beta}+\omega_{i\alpha\beta}\delta_{ij}\right]\delta(g_{i\alpha},g_{j\beta}),
\end{equation}
where $C_{\text{norm}}=\sum_{i=1}^{N}\sum_{\alpha=1}^{\mathcal{L}}\left(\sum_{j=1}^{N}\rho_{ij\alpha}+\sum_{\beta=1}^{\mathcal{L}}\omega_{i\alpha\beta}\right)$.
Parameter $\gamma_{\alpha}$ represents the resolution in layer $\alpha$ \cite{fortunato2007resolution},
and we will discuss the selection of $\gamma_{\alpha}$ in section~\ref{champ}. We remind that $\omega_{i\alpha\beta}$ is the empirical co-expression of gene $i$ between tissues $\alpha$ and $\beta$.
We double-count $(i, j)$ and $(j, i)$, with $i\neq j$, in Eq.~\eqref{eq:corrmultimod} following previous literature \cite{macmahon2015community, bazzi2016community}.

We use a configuration model for correlation matrices \cite{masuda2018configuration} as the null model, while other null models are also possible, such as the H-Q-S algorithm \cite{hirschberger2007randomly} and those derived from random matrix theory \cite{macmahon2015community}.
The configuration model \cite{masuda2018configuration}, implemented in the configcorr package \cite{configcorr-package}, generates the correlation matrix maximizing the entropy under the constraint that the strength (i.e., weighted degree) of each node of the input correlation matrix is conserved. The model assumes normality of the input data. While the algorithm accepts a covariance matrix or a correlation matrix as input, if the input is a covariance matrix, it is first transformed to the correlation matrix before being fed to the configuration model.
To maximize $Q$ given by Eq.~\eqref{eq:corrmultimod}, we feed the supra-modularity matrix $B$, where
$B_{i\alpha j\beta} = (\rho_{ij\alpha}-\gamma_{\alpha}\langle\rho_{ij\alpha}\rangle)\delta_{\alpha\beta}+\omega_{i\alpha\beta}\delta_{ij}$,
to GenLouvain. Again, we use the iterated GenLouvain function \cite{genlouvain} and a consensus clustering technique to obtain a final partition \cite{bassett2013robust} but by inputting $200$ partitions of the same network.


Prior studies developed methods to assess statistical significance of the detected communities in single-layer networks \cite{lancichinetti2010statistical, yang2015defining, kojaku2018generalised}. 
Here, we extend this approach to the case of multilayer correlation matrices and multilayer networks. We do this by comparing a detected community to the same set of nodes in a random graph (or null model) in terms of some quality measure. For each detected community and given quality measure, we calculated the Z score defined by
\begin{equation} \label{eq:zscore}
z=\frac{x-\mu}{\sigma},
\end{equation}
where $x$ is the quality measure calculated for the empirical community, and $\mu$ and $\sigma$ are the expected value and the standard deviation, respectively, of the same quality measure for the same community but under a null model. 
In the following text, we explain this method for multilayer correlation matrices, which we primarily use for our gene data analysis.
We show the details of our methods for general multilayer networks in Text B in \nameref{S1Text}.

We introduce a quality measure of a community that is analogous to the total weight of the intralayer edges within the community. Let $W$ be the total weight of intralayer edges within the set of nodes $S$ in a multilayer correlation matrix. In the remainder of this section, we use the covariance matrices instead of correlation matrices for analytical tractability. This assumption is not detrimental to the application of our methods to multilayer correlation matrix data because a correlation matrix is a covariance matrix in general. Let $C_{ij\alpha}^{\text{org}}$ be the $(i,j)$-th element of $C_{\alpha}^{\text{org}}$, an empirical covariance matrix for layer $\alpha$. Then, we have
\begin{equation} 
\label{eq:totweight}
W = \sum_{\alpha=1}^{\mathcal{L}}\sum_{\substack{i=1 \\ (i,\alpha)\in S}}^{N}\sum_{\substack{j=1\\ (j, \alpha)\in S}}^{i-1} C_{ij\alpha}^{\text{org}},
\end{equation}
where $(i,\alpha)$ represents gene $i$ in layer $\alpha$, and the summation is over all node pairs $((i,\alpha), (j,\alpha))$ in $S$. We exclude the diagonal elements, i.e., $C_{ii\alpha}^{\text{org}}$ in Eq.~\eqref{eq:totweight} because they are equal to 1 for correlation matrices. 

Let $C_{\alpha}^{\text{con}}$ be a sample covariance matrix for layer $\alpha$ generated by the configuration model for correlation matrices \cite{masuda2018configuration}. Let $C_{ij\alpha}^{\text{con}}$ be the $(i,j)$-th element of $C_{\alpha}^{\text{con}}$. Using $\text{E}[C_{\alpha}^{\text{con}}]=C_{\alpha}$, where $C_{\alpha}$ is the covariance matrix for the estimated multivariate normal distribution for layer $\alpha$ \cite{masuda2018configuration}, we obtain
 \begin{equation} \label{eq:Etotweight}
\text{E}\left[\sum_{\alpha=1}^{\mathcal{L}}\sum_{\substack{i=1 \\ (i,\alpha) \in S}}^{N}\sum_{\substack{j=1\\ (j, \alpha)\in S}}^{i-1}C_{ij\alpha}^{\text{con}}\right]=\sum_{\alpha=1}^{\mathcal{L}}\sum_{\substack{i=1 \\ (i,\alpha) \in S}}^{N}\sum_{\substack{j=1\\ (j,\alpha) \in S}}^{i-1} \text{E}[C_{ij\alpha}^{\text{con}}]=\sum_{\alpha=1}^{\mathcal{L}}\sum_{\substack{i=1 \\ (i,\alpha) \in S}}^{N}\sum_{\substack{j=1\\ (j, \alpha) \in S}}^{i-1} C_{ij\alpha}.
\end{equation}
We obtain
\begin{align} \label{eq:Vartotweight0}
\text{Var}\left[\sum_{\alpha=1}^{\mathcal{L}}\sum_{\substack{i=1 \\ (i,\alpha) \in S}}^{N}\sum_{\substack{j=1 \\ (j,\alpha) \in S}}^{i-1}C_{ij\alpha}^{\text{con}}\right] =& \frac{1}{L}\left[\sum_{\alpha=1}^{\mathcal{L}}\sum_{\substack{i=1 \\ (i,\alpha) \in S}}^{N}\sum_{\substack{j=1\\ (j,\alpha) \in S}}^{i-1}\sum_{\substack{k=1 \\ (k,\alpha) \in S}}^{N}\sum_{\substack{r=1 \\ (r,\alpha) \in S}}^{k-1}(C_{ik\alpha}C_{jr\alpha}+C_{ir\alpha}C_{jk\alpha})\right].
\end{align}
We show the derivation of Eq.~\eqref{eq:Vartotweight0} in Text C in \nameref{S1Text}. Note that 
\begin{equation}
\text{Var}\left[\sum_{\alpha=1}^{\mathcal{L}}\sum_{\substack{i=1 \\ (i,\alpha) \in S}}^{N}\sum_{\substack{j=1 \\ (j, \alpha) \in S}}^{i-1}C_{ij\alpha}^{\text{con}}\right]\propto\frac{1}{L},
\end{equation}
which is consistent with the central limit theorem.

\subsection{Determining a resolution parameter value} \label{champ}
For simplicity, we assume $\gamma_{\alpha}$ to be common for all layers and denote the common value by $\gamma$.
We use the Convex Hull of Admissible Modularity Partitions (CHAMP) algorithm version 2.1.0 \cite{weir2017post, WeirCHAMP} to determine the $\gamma$ value. The CHAMP algorithm takes a set of partitions generated by any community detection method as input and identifies the parameter regions in which each partition attains the largest modularity among all the partitions. The algorithm then obtains a pruned subset of admissible partitions and allows one to select parameter values corresponding to more robust community structures, which are large parameter regions in which the same partition maximizes the modularity.

Because we inform the interlayer coupling strength values by the empirical data as we described in section~\ref{multinetworkconstruction},
we only need to tune the $\gamma$ value. Therefore, using $15$ evenly spaced $\gamma$ values ranging from $\gamma=1$ to $\gamma=4$, we run a multilayer community detection method to obtain $15$ partitions, one for each $\gamma$ value, for a given multilayer network.
%
%
Then, we employ the one-dimensional CHAMP on the $15$ corresponding partitions to identify the ranges of $\gamma$ in which the same partition maximizes the modularity. The wider ranges of $\gamma$ correspond to more robust ranges of $\gamma$, so we choose a $\gamma$ value in the two widest ranges according to CHAMP.

\subsection{Specialist and generalist communities}
The communities in multilayer correlation matrices and multilayer networks determined by the maximization of multilayer modularity may span multiple layers. We refer to a community containing genes belonging to various layers, i.e., tissues, as a generalist community. We refer to a community that contains genes in mostly just one tissue as a specialist community. The genes in a generalist community are general in the sense that they are co-expressed similarly across multiple tissues, whereas the genes in a specialist community are specialist in the sense that they are uniquely co-expressed in a single tissue. We will give the precise definitions of a generalist community and a specialist community in the following text. These different types of communities occur due to the similarity or difference between gene co-expression patterns across different tissues. In particular, some pairs of genes show co-expression across individuals in only specific tissues and others in multiple tissues. We are interested in whether our community detection method can detect these different types of communities. Therefore, we need a measure to classify each detected community as a generalist community or a specialist community.

We define a measure called the specialist fraction to quantify how specialized any multilayer community is as follows.
For a given community, we first find the number of genes unique to each tissue $\alpha$, i.e., the genes $i$ for which node ($i$, $\alpha$) belongs to the community and node ($i$, $\beta$) does not for any $\beta \neq \alpha$.
Second, we define the specialist tissue of the community as the tissue that has the largest number of unique genes.
The specialist fraction is the number of genes unique to the specialist tissue divided by the total number of nodes in the community.
If the community lies within one layer, the specialist fraction is equal to $1$.
A large value of the specialist fraction suggests that the community is a specialist community. Genes unique to a specialist community may have functions specific to the tissue. In contrast, if all genes belong to at least two tissues, the specialist fraction is equal to $0$.
If many genes belong to different tissues in the community, the specialist fraction is low, suggesting that the community is relatively a generalist community. Genes in a generalist community may have functions expressed across various tissues.

\subsection{Gene set enrichment analysis} \label{enrichmentanalysis}
To explore the biological processes associated with the set of genes constituting a detected community, we carried out a gene set enrichment analysis. 
It is a standard method for detecting statistically significant enriched biological processes, pathways, regulatory motifs, protein complexes, and disease phenotypes in the given gene set. We use g:Profiler (version e109\textunderscore eg56\textunderscore p17\textunderscore 1d3191d) for this purpose \cite{raudvere2019g} and restrict our analysis to the Gene Ontology biological process (GO:BP) release 2023-03-06 \cite{ashburner2000gene, gene2023gene} and Human phenotype ontology (HP) release 2023-01-27 \cite{kohler2021human} results. We use a Benjamini-Hochberg FDR significance threshold \cite{benjamini1995controlling} of 0.05.

\subsection{Localization of genes on chromosomes} \label{localization}
We developed statistical methods to investigate whether the genes in a community detected by our community detection method are physically clustered across the genome. To this end, we first ask whether a group of genes are more frequently located on the same chromosome than a control.
Consider a group of genes, denoted by $c$. Let $n$ be the number of genes in group $c$. We define the fraction of pairs of genes on the same chromosome as 
\begin{equation}
\label{eq:chrom_score}
x_{c}=\frac{\text{number of pairs of genes in group } c \text{ on the same chromosome}} {n(n-1)/2}.
\end{equation}
The denominator of $x_{c}$ is equal to the number of pairs of genes in group $c$ and gives the normalization. For the control, we uniformly randomly shuffle the association between the $N=203$ genes that we initially selected for our analysis and the chromosome to which each of the $N$ genes belongs. After this random shuffling, the $n$ genes are randomly distributed on various chromosomes as the $N=203$ genes are distributed on those chromosomes. Then, we calculate $x_{c}^{\text{rand}}$ according to this random distribution of the $n$ genes using Eq.~\eqref{eq:chrom_score}. We repeat this randomization $100$ times and calculate the average and standard deviation of $x_{c}^{\text{rand}}$, and then the Z score. If the Z score is significantly positive, then we say that the group of genes $c$ has more pairs of genes on the same chromosome than the control. 

Second, we tested whether the genes in $c$ are located closer to each other on the chromosome than a control, 
given the number of genes in $c$ on each chromosome. 
To this end, we define the physical distance measured in base pairs between gene $i$ and gene $j$ on the same chromosome, $d(i, j)$, as follows. Without loss of generality, assume that the end position of gene $i$ is less than the start position of gene $j$. Then, we set
\begin{equation} \label{eq:dist}
d(i,j)=(\text{start position of gene $j$}) - (\text{end position of gene $i$}).
\end{equation}
Furthermore, we define the average distance between genes in group $c$ as 
\begin{equation} \label{eq:avg_dist}
d_{c}=\frac{\sum_{i,j\text{ in group $c$ on the same chromosome}}d(i, j)}{\text{number of pairs of genes in group $c$ on the same chromosome}}.
\end{equation}
Denote by $n_k$ the number of genes in group $c$ that are on chromosome $k$.
Note that the denominator in Eq.~\eqref{eq:avg_dist} is equal to $\sum_k n_k(n_k-1)/2$.
For the control, for each $k$, we choose $n_k$ genes uniformly at random out of all genes on chromosome $k$ from the $N$ genes. We carry out this procedure for all chromosomes $k$ on which there are at least two genes in group $c$ (i.e., $n_k \ge 2$). Then, we calculate $d_c$ for this random distribution of genes, which we refer to as $d_{c}^{\text{rand}}$. We repeat this randomization $100$ times and calculate the average and standard deviation of $d_{c}^{\text{rand}}$, and then the Z score. If the Z score is significantly negative, then we say that the genes in group $c$ are localized on the chromosomes.

Third, we test whether the genes in $c$ are located closer to each other than a control on a given chromosome. We define the average distance between genes in group $c$ on chromosome $k$ as
\begin{equation} \label{eq:avg_dist_chrom}
\tilde{d}_{c,k}=\frac{\sum_{i,j\text{ in group $c$ on chromosome $k$}}d(i, j)}{ n_k(n_k-1)/2 }.
\end{equation}
For the control, we choose $n_k$ genes uniformly at random out of all genes that are among the $N$ genes and on chromosome $k$.
Then, we calculate $\tilde{d}_{c,k}$ for this random distribution of genes, which we refer to as $\tilde{d}_{c,k}^{\text{rand}}$.
We repeat this randomization $100$ times and calculate the average and standard deviation of $\tilde{d}_{c,k}^{\text{rand}}$, and then the Z score.
We carry out this procedure for each chromosome $k$ on which there are at least two genes in group $c$ (i.e., $n_k \ge 2$). We apply the Bonferroni correction \cite{dunn1961multiple} separately to each $c$ to determine which communities have a significantly smaller average distance between pairs of genes on a specific chromosome than the control. We chose to apply the Bonferroni correction because it is a more conservative statistical method than others, such as FDR.

\subsection{Pancreas-specific cis-eQTL analysis} \label{eqtl_methods}
Expression quantitative trait loci (eQTL) analysis identifies variants that have significant associations with expression levels of specific genes. We hypothesize that changes in expression levels of a pair of co-expressed genes are associated with the same set of variants. If true, we expect to identify variants that are associated with the expression of both genes in the pair. To investigate gene pairs with shared eQTL single nucleotide polymorphisms (SNPs) in the pancreas, we downloaded the cis-eQTL data set from GTEx release V8. This data set involves SNP-gene pairs with association significance indicated with a nominal $p$ value. The changes in the expression levels of a given gene may be associated with one or multiple SNPs. Alternatively, it may have no eQTLs, meaning that no SNPs are associated with its gene expression. Using this data set, we searched for SNPs that were associated with both of the genes in a given gene pair of interest. Given that we are interested in whether co-expressed genes share common SNPs, we only investigate gene pairs with co-expression (as defined by Eq.~\eqref{eq:genecorr}) greater than $0.5$ in the pancreas.

\section{Results} \label{results}
\subsection{Communities in the multilayer correlation matrix} \label{part_corr_results}
We compare the gene communities obtained from the multilayer correlation matrix and those obtained from multilayer gene networks constructed using graphical lasso. For a brief review of graphical lasso, see Text D in \nameref{S1Text}.

\begin{figure}[!h]
\centering
\includegraphics[width=\linewidth]{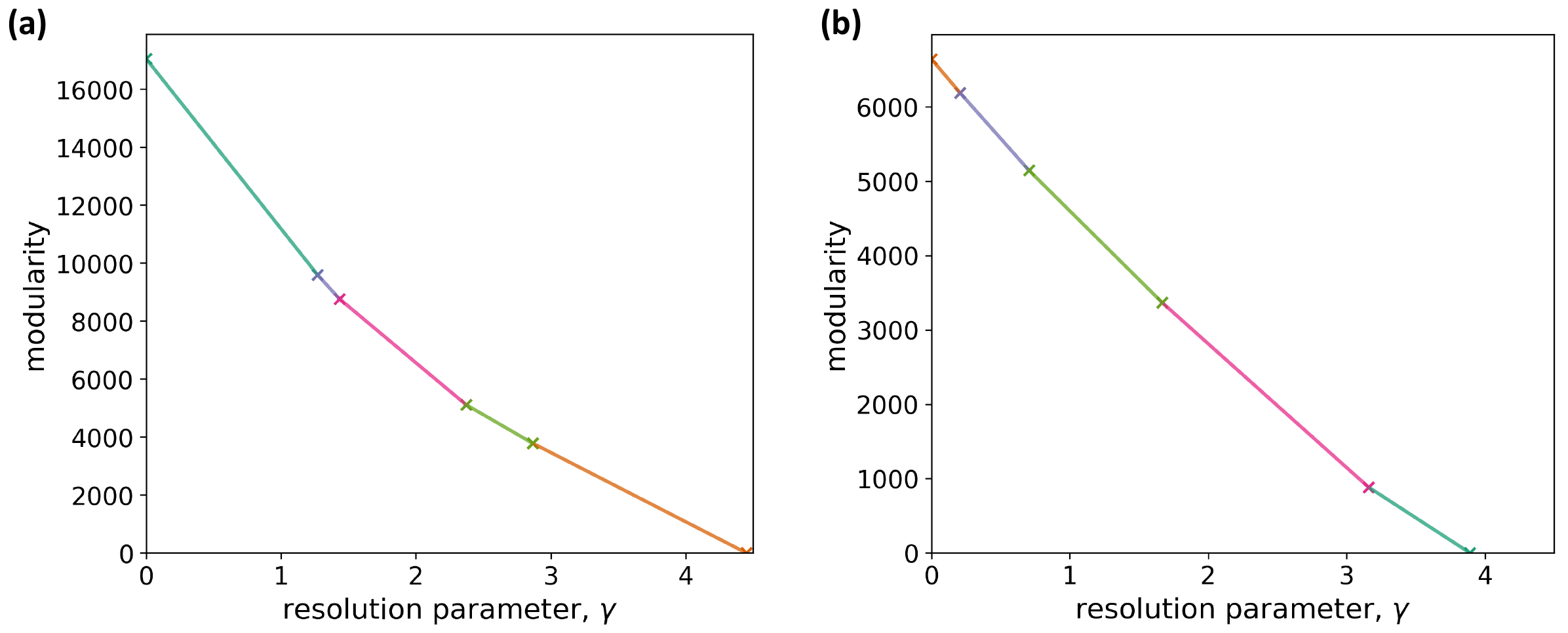}
\caption{{\bf Determination of the resolution parameter value by CHAMP.}
(a) Multilayer correlation matrix. (b) Multilayer network obtained by graphical lasso. The convex hull of the lines in the $(\gamma, Q)$ plane, each of which corresponds to a node partitioning, is a piecewise linear curve with the transition values indicated by a cross and change in the line color. Each line segment corresponds to the optimal node partitioning in the corresponding range of $\gamma$.}
\label{fig:CHAMP}
\end{figure}

We run iterated GenLouvain \cite{genlouvain} on the multilayer correlation matrix to approximately maximize the multilayer modularity at each value of the resolution parameter, $\gamma$, which we assume to be common for all layers. We then use the CHAMP algorithm to determine optimal values of $\gamma$~\cite{weir2017post, WeirCHAMP}. We show the results of CHAMP in Fig~\ref{fig:CHAMP}(a). The figure indicates that robust ranges of $\gamma$, which are relatively wide ranges of $\gamma$ in which the optimal partition is the same and correspond to relatively long straight line segments in the figure, are approximately $0<\gamma<1.2$ or $2.9<\gamma<4.4$. Therefore, we examine the node partitions with one arbitrary $\gamma$ value from each of these two stable regions of $\gamma$, i.e., $\gamma=1$ and $\gamma=3$.

\begin{figure}[!h]
\centering
\includegraphics[width=\linewidth]{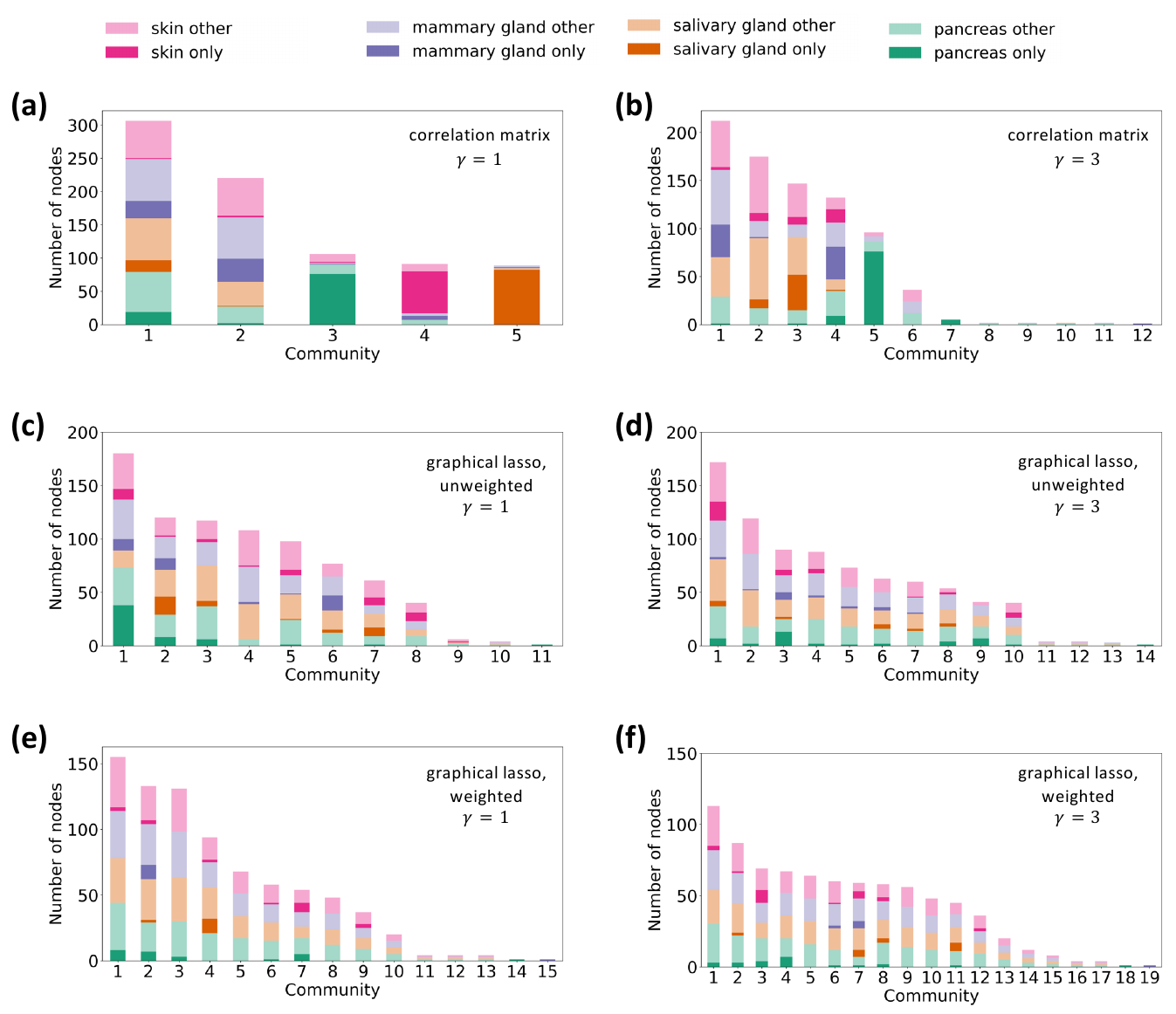}
\caption{{\bf Composition of each community by layer, i.e., tissue.}
(a) Multilayer correlation matrix, $\gamma=1$. (b) Multilayer correlation matrix, $\gamma=3$. (c) Unweighted multilayer network obtained by graphical lasso, $\gamma=1$. (d) Unweighted multilayer network obtained by graphical lasso, $\gamma=3$. (e) Weighted multilayer network obtained by graphical lasso, $\gamma=1$. (f) Weighted multilayer network obtained by graphical lasso, $\gamma=3$. The darker shades indicate nodes corresponding to genes that only appear in one layer in the given community. The lighter shades indicate nodes corresponding to genes that appear in multiple layers in the community.}
\label{fig:barcharts}
\end{figure}

We show the composition of the resulting node partitions with $\gamma=1$ and $\gamma=3$ in Fig~\ref{fig:barcharts}(a) and \ref{fig:barcharts}(b), respectively. As expected, the number of communities increases when $\gamma$ increases. 
We show the Z score for the total intralayer weight within each community detected with $\gamma=1$ and $\gamma=3$ in Table~\ref{table:corr_intra_zscores}. With $\gamma=3$, communities 8 through 12 contain no intralayer edges such that one cannot run the randomization, leading to a null Z score. 
These communities contain only one gene; communities 8, 9, 10, and 11 detected with $\gamma=3$ contain two nodes representing the same gene in two different tissues, and community 12 contains only one node. We omitted these trivial communities in Table~\ref{table:corr_intra_zscores}. The table indicates that all the communities detected with $\gamma=1$ and all the communities containing at least two genes detected with $\gamma=3$ (i.e., communities 1 through 7) are statistically significant.  

\begin{table}[!ht]
\centering
\caption{
{\bf Z scores for the total intralayer weight within each community detected in the multilayer correlation matrix.}}
\begin{tabular}[b]{|c c|c c|}\hline
\multicolumn{2}{|c|}{$\gamma=1$} & \multicolumn{2}{c|}{$\gamma=3$} \\\hline
Comm.       & Z score      & Comm.       & Z score      \\\hline
1               &     $24.432$      & 1               &    $68.526$          \\
2               &     $73.282$      & 2               &     $55.267$        \\
3               &     $62.569$      & 3               &      $72.008$        \\
4               &     $14.972$         & 4               &       $19.071$       \\
5               &     $65.318$         & 5               &     $124.080$        \\
                &              & 6               &       $40.288$     \\
                &              & 7               &       $14.699$       \\\hline
\end{tabular}
\begin{flushleft} Comm.\,denotes community.
\end{flushleft}
\label{table:corr_intra_zscores}
\end{table}

Both node partitions contain some communities that appear to be generalist communities and other communities that appear to be specialist communities. We remind that a generalist community indicates genes that are similarly co-expressed in multiple tissues and that a specialist community indicates genes that are uniquely co-expressed in one tissue.
To quantify these findings, we show in Table~\ref{table:specialist_frac} the specialist fraction for each community in the partition with $\gamma=1$. Communities 3, 4, and 5 have specialist fractions greater than $0.5$, so we regard them as specialist communities. In contrast, because communities 1 and 2 have specialist fractions substantially less than $0.5$, we regard them as generalist communities.
The same table also shows the specialist fraction for each significant community found with $\gamma=3$. Communities 5 and 7 have specialist fractions greater than $0.5$. Both of them are pancreas specialist communities. We regard communities 1, 2, 3, 4, and 6, whose specialist fraction is substantially less than $0.5$, as generalist communities.

\begin{table}[!ht]
\centering
\caption{
{\bf Specialist fraction and the corresponding tissue for each community detected in the multilayer correlation matrix and for each community detected in the multilayer networks obtained by graphical lasso.}}
\begin{tabular}[b]{|c c c c c | c c c c c|}\hline
\multicolumn{10}{|c|}{Multilayer correlation matrix} \\\hline
  \multicolumn{5}{|c|}{$\gamma=1$} & \multicolumn{5}{c|}{$\gamma=3$}\\\hline
  Comm. & \thead{No. \\  genes} & \thead{No. \\ specialist \\ genes} & \thead{Specialist \\ fraction} & \thead{Specialist \\ tissue} & Comm. & \thead{No. \\  genes} & \thead{No. \\ specialist \\ genes} & \thead{Specialist \\ fraction} & \thead{Specialist \\ tissue} \\ \hline
  1 & $153$ & $26$ & $0.085$ & mammary gland & 1 & $96$ & $34$ & $0.160$ & mammary gland\\
  2 & $104$ & $35$ & $0.159$ & mammary gland & 2 & $84$ & $9$ & $0.051$ & salivary gland\\
  3 & $92$ & $76$ & $0.717$ & pancreas & 3 & $87$ & $37$ & $0.252$ & salivary gland\\
  4 & $80$ & $63$ & $0.692$ & skin & 4 & $88$ & $34$ & $0.258$ & mammary gland\\
  5 & $86$ & $82$ & $0.921$ & salivary gland & 5 & $86$ & $76$ & $0.792$ & pancreas\\
   & & & & & 6 & 12 & 0 & 0.000 & N/A\\
   & & & & & 7 & $5$ & $5$ & $1.000$ & pancreas\\\hline \hline
  
 \multicolumn{10}{|c|}{Unweighted multilayer network obtained by graphical lasso} \\\hline
  \multicolumn{5}{|c|}{$\gamma=1$} & \multicolumn{5}{c|}{$\gamma=3$}\\\hline
  Comm. & \thead{No. \\  genes} & \thead{No. \\ specialist \\ genes} & \thead{Specialist \\ fraction} & \thead{Specialist \\ tissue} & Comm. & \thead{No. \\  genes} & \thead{No. \\ specialist \\ genes} & \thead{Specialist \\ fraction} & \thead{Specialist \\ tissue} \\ \hline
  1 & $102$ & $38$ & $0.211$ & pancreas & 1 & $80$ & $18$ & $0.105$ & skin\\ 
  2 & $62$ & $17$ & $0.142$ & salivary gland & 2 & $37$ & $2$ & $0.017$ & pancreas\\ 
  3 & $48$ & $6$ & $0.051$ & pancreas & 3 & $47$ & $13$ & $0.144$ & pancreas\\ 
  4 & $36$ & $2$ & $0.019$ & mammary gland & 4 & $33$ & $4$ & $0.045$ & skin\\ 
  5 & $35$ & $5$ & $0.051$ & skin & 5 & $21$ & $2$ & $0.027$ & mammary gland\\
  6 & $35$ & $14$ & $0.182$ & mammary gland & 6 & $24$ & $4$ & $0.063$ & salivary gland\\
  7 & $32$ & $8$ & $0.131$ & salivary gland & 7 & $18$ & $2$ & $0.033$ & salivary gland\\
  8 & $17$ & $8$ & $0.200$ & skin & 8 & $23$ & $4$ & $0.074$ & pancreas\\
  9 & $3$ & $1$ & $0.167$ & skin & 9 & $18$ & $7$ & $0.171$ & pancreas\\
   & & & & & 10 & $15$ & $5$ & $0.125$ & skin\\\hline \hline
   
   \multicolumn{10}{|c|}{Weighted multilayer network obtained by graphical lasso} \\\hline
  \multicolumn{5}{|c|}{$\gamma=1$} & \multicolumn{5}{c|}{$\gamma=3$}\\\hline
  Comm. & \thead{No. \\  genes} & \thead{No. \\ specialist \\ genes} & \thead{Specialist \\ fraction} & \thead{Specialist \\ tissue} & Comm. & \thead{No. \\  genes} & \thead{No. \\ specialist \\ genes} & \thead{Specialist \\ fraction} & \thead{Specialist \\ tissue} \\ \hline 
  1 & $51$ & $8$ & $0.052$ & pancreas & 1 & $35$ & $3$ & $0.027$ & pancreas\\
  2 & $54$ & $11$ & $0.083$ & mammary gland & 2 & $30$ & $3$ & $0.034$ & pancreas\\
  3 & $38$ & $3$ & $0.023$ & pancreas & 3 & $31$ & $9$ & $0.130$ & skin\\
  4 & $38$ & $11$ & $0.117$ & salivary gland & 4 & $23$ & $7$ & $0.104$ & pancreas\\
  5 & $17$ & $0$ & $0.000$ & N/A & 5 & $16$ & $0$ & $0.000$ & N/A\\
  6 & $16$ & $1$ & $0.017$ & pancreas & 6 & $19$ & $2$ & $0.033$ & mammary gland\\
  7 & $24$ & $7$ & $0.130$ & skin & 7 & $32$ & $5$ & $0.085$ & salivary gland\\
  8 & $12$ & $0$ & $0.000$ & N/A & 8 & $23$ & $3$ & $0.052$ & salivary gland\\
  9 & $12$ & $3$ & $0.081$ & skin & 9 & $14$ & $0$ & $0.000$ & N/A\\
  10 & $5$ & $0$ & $0.000$ & N/A & 10 & $12$ & $0$ & $0.000$ & N/A\\
   & & & & & 11 & $19$ & $6$ & $0.133$ & salivary gland\\
   & & & & & 12 & $11$ & $2$ & $0.056$ & skin\\
   & & & & & 13 & $5$ & $0$ & $0.000$ & N/A\\
   & & & & & 14 & $3$ & $0$ & $0.000$ & N/A\\
   & & & & & 15 & $2$ & $0$ & $0.000$ & N/A\\\hline
  
\end{tabular}
\begin{flushleft} Comm.\,denotes community and No.\,denotes ``number of''.
\end{flushleft}
\label{table:specialist_frac}
\end{table}

\subsection{Communities in the multilayer networks obtained by graphical lasso} \label{part_lasso_results}
For comparison purposes, we run the iterated GenLouvain on the multilayer networks that we constructed using graphical lasso (see Text D in \nameref{S1Text} for the methods). The results of CHAMP on the detected node partition of the unweighted network, shown in Fig~\ref{fig:CHAMP}(b), indicate that the optimal ranges of $\gamma$ are approximately $0.7<\gamma<1.7$ or $1.7<\gamma<3.2$. Therefore, we use the same $\gamma$ values as those for our multilayer correlation matrix, i.e., $\gamma=1$ and $\gamma=3$.

We show the composition of the resulting node partitions of the unweighted network obtained using graphical lasso with $\gamma=1$ and $\gamma=3$ in Fig~\ref{fig:barcharts}(c) and \ref{fig:barcharts}(d), respectively. With $\gamma=1$, we find eleven communities, nine of which are significant. With $\gamma=3$, we find fourteen communities, ten of which are significant.
See Text B in \nameref{S1Text} for the statistical results. We also show the composition of the node partitions of the weighted multilayer network obtained using graphical lasso with $\gamma=1$ and $\gamma=3$ in Fig~\ref{fig:barcharts}(e) and \ref{fig:barcharts}(f), respectively.

Fig~\ref{fig:barcharts}(c)--(f) suggests that these partitions apparently contain generalist communities only.
Table~\ref{table:specialist_frac} shows the specialist fraction for each significant community in the unweighted network and each community in the weighted network. 
Note that we have not evaluated the significance of the communities detected for the weighted multilayer network because the configuration model for weighted networks, which is necessary for constructing a significance test, is not a straightforward concept \cite{squartini2011analytical, mastrandrea2014enhanced}.
For the unweighted network, with both $\gamma=1$ and $\gamma=3$, all the significant communities have specialist fractions at most $0.211$. For the weighted network, with both $\gamma=1$ and $\gamma=3$, all the communities with more than one gene have specialist fractions at most $0.133$. Therefore, we conclude that there are no specialist communities for either the unweighted or weighted network and with either $\gamma=1$ or $\gamma=3$. 

In sum, our community detection method on correlation matrices finds tissue-specific gene co-expression patterns, evident by the detection of specialist communities, whereas the graphical lasso does not. Because we are interested in comparing the biological implications of specialist communities versus generalist communities, in the following sections,
we only analyze the communities detected for our multilayer correlation matrix.
In particular, we will carry out tissue-specific analysis to investigate the specialist communities detected by our method.

\subsection{Localization of genes on chromosomes} \label{init_local_results}
To investigate the possible localization of genes in the detected communities on the chromosomes, we first analyze whether the $N=203$ among the $56,200$ genes that we are analyzing in the GTEx data set are already localized in the genome. The Z score for a fraction of pairs of genes on the same chromosome is 6.735 ($p <10^{-6}$), which suggests that the $N=203$ genes are distributed on different chromosomes in a highly biased manner relative to how all the $56,200$ genes are distributed. The Z score for the average distance
between pairs of genes on the same chromosome is $-6.059$ ($p <10^{-6}$). Therefore, the average distance between pairs of genes among the $N=203$ genes is significantly smaller than by chance. This result is expected given that highly expressed genes in glandular tissues cluster in specific loci \cite{saitou2020functional}. We show the Z scores for the average distance between pairs of genes on each chromosome, analyzed separately, in Table \ref{table:zscores_avg_dist_allgenes}. At a significance level of $p=0.05$, there is significant localization of genes on chromosomes 2 ($p=0.0088$; Bonferroni corrected; same for the following $p$ values), 4 ($p<10^{-4}$), 12 ($p=0.0098$), and 17 ($p<10^{-4}$).

\begin{table}[!ht]
\centering
\caption{
{\bf Z score for the average distance between pairs of genes on each chromosome for the $N=203$ genes.}}
\begin{tabular}[b]{|c c|c c|}\hline
  Chr & Z score & Chr & Z score\\ \hline
  1 & $-1.881$ & 14 & $0.085$\\
  2 & $-3.540$ & 15 & $-0.404$\\
  3 & $1.293$ & 16 & $0.599$\\
  4 & $-4.736$ & 17 & $-4.842$\\
  5 & $-1.858$ & 18 & N/A\\
  6 & $-0.422$ & 19 & $-2.545$\\
  7 & $-0.873$ & 20 & $-1.458$\\
  8 & $0.691$ & 21 & $-0.317$\\
  9 & $0.276$ & 22 & $-0.564$\\
  10 & $0.552$ & X & $2.103$\\
  11 & $-1.112$ & Y & N/A\\
  12 & $-3.512$ & M & $-0.858$\\
  13 & N/A & & \\\hline
\end{tabular}
\begin{flushleft} M stands for the mitochondrial chromosome. Chr denotes chromosome.
\end{flushleft}
\label{table:zscores_avg_dist_allgenes}
\end{table}

Next, we run the same localization analysis for each community in the multilayer correlation matrix detected with $\gamma=1$ and $\gamma=3$. For a generalist community, we only included the genes in the community that appear in at least three out of the four tissues in this analysis. This is because such genes may play functional roles, which the generalist community represents, across many types of tissues. With this restriction, each gene is present in at most one generalist community. Note that, without this restriction, a gene may appear in multiple generalist communities because the four nodes in the multilayer network representing the same gene may belong to different communities. We exclude this case for simplicity.

For each community, we show in Table \ref{table:loc_zscores} the Z score for the fraction of pairs of genes in the community that are on the same chromosome. With $\gamma=1$, communities 2 ($p<10^{-4}$) and 3 ($p=4.05\cdot 10^{-4}$) have significantly more genes among the $N=203$ genes on the same chromosome than by chance. The same table also shows the Z score for the average distance on the chromosome between pairs of genes in the same community for each community. We find that, with $\gamma=1$, community 2 has a significantly smaller average gene-to-gene distance than by chance ($p=0.0336$).
With $\gamma=3$, communities 1 ($p<10^{-4}$), 5 ($p<10^{-4}$), and 6 ($p<10^{-4}$) have significantly more pairs of genes on the same chromosome than by chance, and community 1 ($p=0.0069$) has a significantly smaller average gene-to-gene distance than by chance (see Table \ref{table:loc_zscores}).

\begin{table}[!ht]
\centering
\caption{
{\bf Analysis of localization of genes in each community detected in the multilayer correlation matrix.}}
\begin{tabular}[b]{|c c c|c c c|}\hline
\multicolumn{3}{|c|}{$\gamma=1$} & \multicolumn{3}{c|}{$\gamma=3$} \\\hline
Comm. & \thead{Z score\\ for $x_c$} & \thead{Z score\\ for $d_c$} & Comm. & \thead{Z score\\ for $x_c$} & \thead{Z score\\ for $d_c$}\\ \hline
1 & $1.520$ & $-1.095$ & 1 & $6.190$ & $-3.251$\\
2 & $7.094$ & $-2.710$ & 2 & $-0.388$ & $0.652$\\
3 & $3.940$ & $-1.245$ & 3 & $0.879$ & $-0.405$\\
4 & $0.160$ & $0.179$ & 4 & $-0.281$ & $0.924$\\
5 & $-0.102$ & $-0.344$ & 5 & $4.485$ & $-1.175$\\
 & & & 6 & $6.634$ & $-2.255$\\
 & & & 7 & $-0.845$ & N/A\\\hline
\end{tabular}
\begin{flushleft} Note that $x_c$ is the normalized fraction of pairs of genes in the community on the same chromosome and that $d_c$ is the normalized distance between two genes in the community on the same chromosome. Comm.\,denotes community.
\end{flushleft}
\label{table:loc_zscores}
\end{table}

We then compute the Z score for the average distance between pairs of genes separately for each chromosome in addition to each community. We exclude the community-chromosome pairs that have less than three genes from this analysis. With both $\gamma=1$ and $\gamma=3$, no group of genes on a specific chromosome in a specific community is significantly clustered when we impose the Bonferroni correction over all the community-chromosome pairs ($45$ and $23$ pairs with $\gamma=1$ and $\gamma=3$, respectively; see Tables B and C in \nameref{S1Text} for the Z scores).
With the Bonferroni correction applied to each community separately, there are still no significant clusters in the partition with $\gamma=1$.
However, with $\gamma=3$, we find that the genes in community 1 on chromosome 1 ($p=0.0199$) and those in community 5 on chromosome 11 ($p=0.0371$) are significantly clustered.

\subsection{Functional analysis of selected communities} \label{local_results}
For the communities detected for our multilayer correlation matrix, we found clusters of physically localized genes within two communities with $\gamma=3$ but none with $\gamma=1$. Because we are interested in exploring biological implications of localized clusters of genes, we carry out further analysis on the node partition with $\gamma=3$ in this section. A table showing which nodes (i.e., genes) belong to which communities in this partition is available on GitHub \cite{githubrepo}.

First, we conducted an enrichment analysis of the communities identified with $\gamma=3$. We started with an enrichment analysis for the top 50 genes that have the highest expression out of the $203$ genes in the network in each tissue. We find that, in all tissues, the top 50 highly expressed genes are enriched significantly in well-established housekeeping categories, such as oxidative phosphorylation and aerobic electron transport chain (FDR $<0.05$; see Table D in \nameref{S1Text}). Echoing this finding, one of the modules that we identified (community 1) shows similar enrichment for mitochondrial function, such as aerobic electron transport chain ($p=1.05\cdot 10^{-10}$) and oxidative phosphorylation ($p=5.90\cdot 10^{-11}$) (see Table E in \nameref{S1Text}). However, in the other six communities, our network approach identifies novel gene modules with functional enrichments in epidermis development (community 2, $p=1.90\cdot 10^{-24}$), keratinization (community 5, $p=1.95\cdot 10^{-19}$), positive regulation of respiratory burst (community 6, $p=5.36\cdot 10^{-8}$), and adaptive thermogenesis (community 7, $p=1.73\cdot 10^{-2}$). Furthermore, these modules are enriched with diseases relevant to the tissues examined, such as hyperkeratosis (community 2, $p=3.05\cdot 10^{-7}$) and recurrent pancreatitis (community 1, $p=1.78\cdot 10^{-19}$). In addition, we analyzed the top 50 highly connected genes (i.e., top 50 genes in terms of the weighted degree, or in other words, top 50 hub genes) in each of the single-layered networks for each tissue. Not surprisingly, this analysis identified genes that are enriched for functions and diseases that are specific to each tissue (see Table F in \nameref{S1Text}). However, we found that most of the genes that are identified in our multilayer network approach are different from those identified with single-layer analysis (see Text G in \nameref{S1Text}). We also found that the functional enrichments of these two network approaches were different (see Table E versus Table F in \nameref{S1Text}). Overall, our method provides additional biological insights than simple expression-level filtering and single-layer network analysis.

\begin{figure}[!h]
\centering
\includegraphics[width=\linewidth]{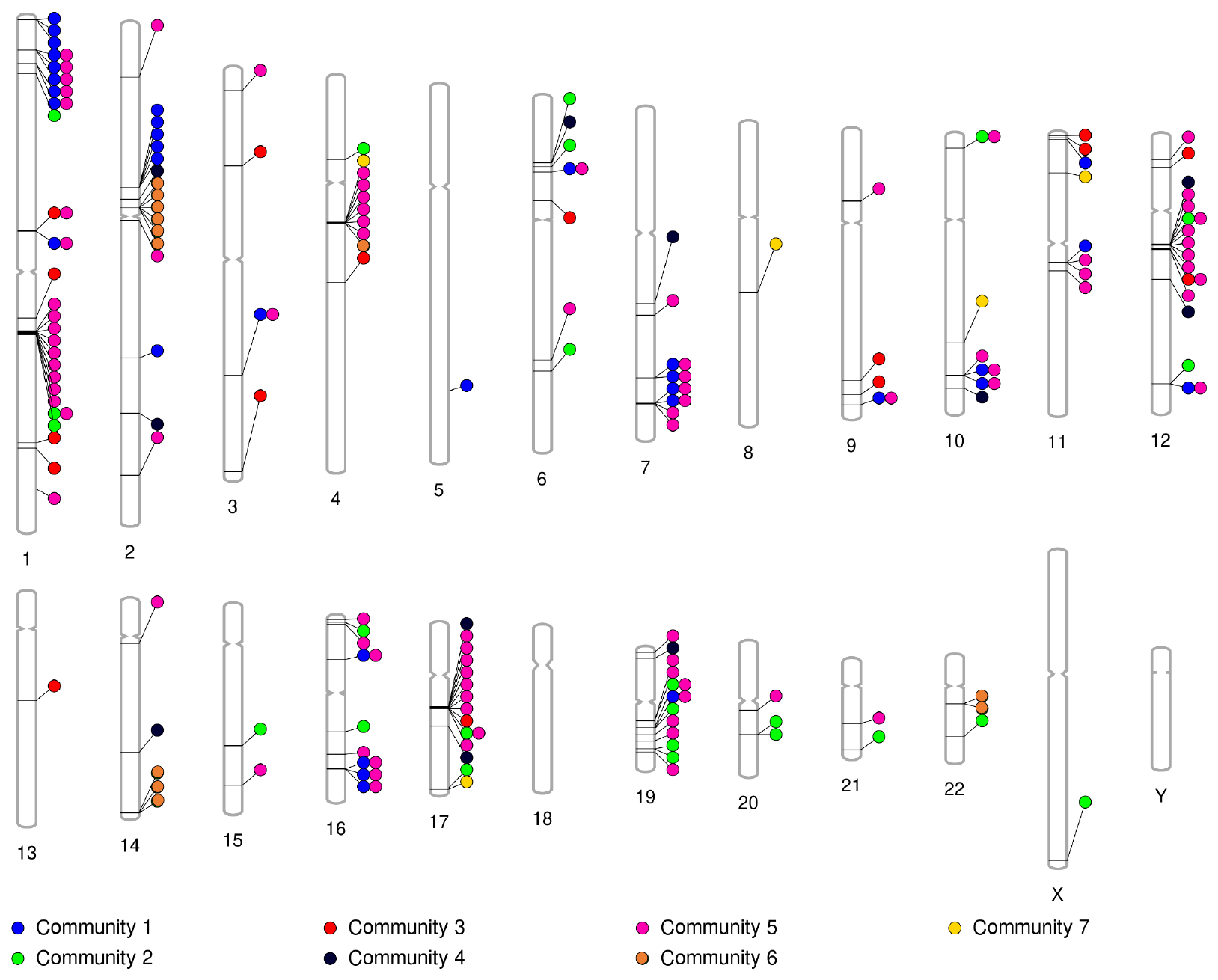}
\caption{{\bf Location of genes on chromosomes, colored by community.} There is a colored circle for each associated community pointing to each gene. Note that a gene can belong to more than one community, denoted by multiple colored circles next to each other horizontally pointing to the same gene. This figure allows us to visually see clusters of genes on specific chromosomes and their associated community.}
\label{fig:chroms}
\end{figure}

Our multilayer network analysis allowed us to investigate genes that are co-expressed in multiple tissues. We surmised that membership of genes in the same community can be facilitated by shared regulatory sequences affecting multiple genes at the same time. Given that regulatory regions affect gene expression in cis (i.e., nearby regions), we hypothesize that genes in the same multilayer community may be physically close to each other. To investigate this, we visualize in Fig~\ref{fig:chroms} the location of the genes in the different communities on the chromosomes. As in the localization analysis presented in section~\ref{init_local_results}, for a generalist community, we only show in Fig~\ref{fig:chroms} the genes in the community that are present in at least three tissues. In Fig~\ref{fig:chroms}, a color of the circles represents a community. Note that a gene can belong to more than one community, denoted by multiple colored circles next to each other horizontally pointing to the same gene. It happens to be the case that a gene is associated with a maximum of two different communities, hence a maximum of two colored circles pointing to the same gene. Visually, Fig~\ref{fig:chroms} suggests some tight clusters of genes, especially in community 5.

In section~\ref{init_local_results}, we found significantly localized clusters of genes in community 1 on chromosome 1 and in community 5 on chromosome 11 in the partition with $\gamma=3$. It is somewhat surprising that only these two community-chromosome pair gene sets are significantly localized because there appear to be more localized clusters in Fig~\ref{fig:chroms}. A possible reason for this discrepancy is that, besides the genes in the community-chromosome pair of interest, there are so few other genes on the chromosome that the random shuffling of gene associations does not provide sufficient randomization. In this case, the empirical average distance between genes in the community-chromosome pair will not be statistically different from the average distance for the randomized data. Therefore, here we directly compared the average distance between pairs of genes on each community-chromosome pair, as defined by Eq.~\eqref{eq:avg_dist_chrom}, to that for community 5 on chromosome 11. We decided to analyze community 5 because it is a pancreas specialist community while community 1 is a generalist community, as we discussed in regards to functional enrichment earlier in this section.

We denote the average distance between the pairs of genes among the three genes in community 5 on chromosome 11 by $\tilde{d}_{5,11}$, calculated using Eq.~\eqref{eq:avg_dist_chrom}. 
We looked for any community-chromosome pair, containing all the genes in the selected community on the selected chromosome, with at least three genes whose average distance between genes is less than $\tilde{d}_{5,11}$.
There are five such additional gene clusters: community 1 on chromosome M, which contains 15 genes, community 5 on chromosome 4, which contains 6 genes, community 5 on chromosome 17, which contains 9 genes, community 6 on chromosome 2, which contains 6 genes, and community 6 on chromosome 14, which contains 3 genes. Among all these community-chromosome pairs, we focused on the three gene clusters in community 5, including the gene cluster on chromosome 11. We opted to do so because community 5 is a pancreas specialist community, whereas communities 1 and 6 are generalist communities. 

\begin{figure}[!h]
\centering
\includegraphics[width=\linewidth]{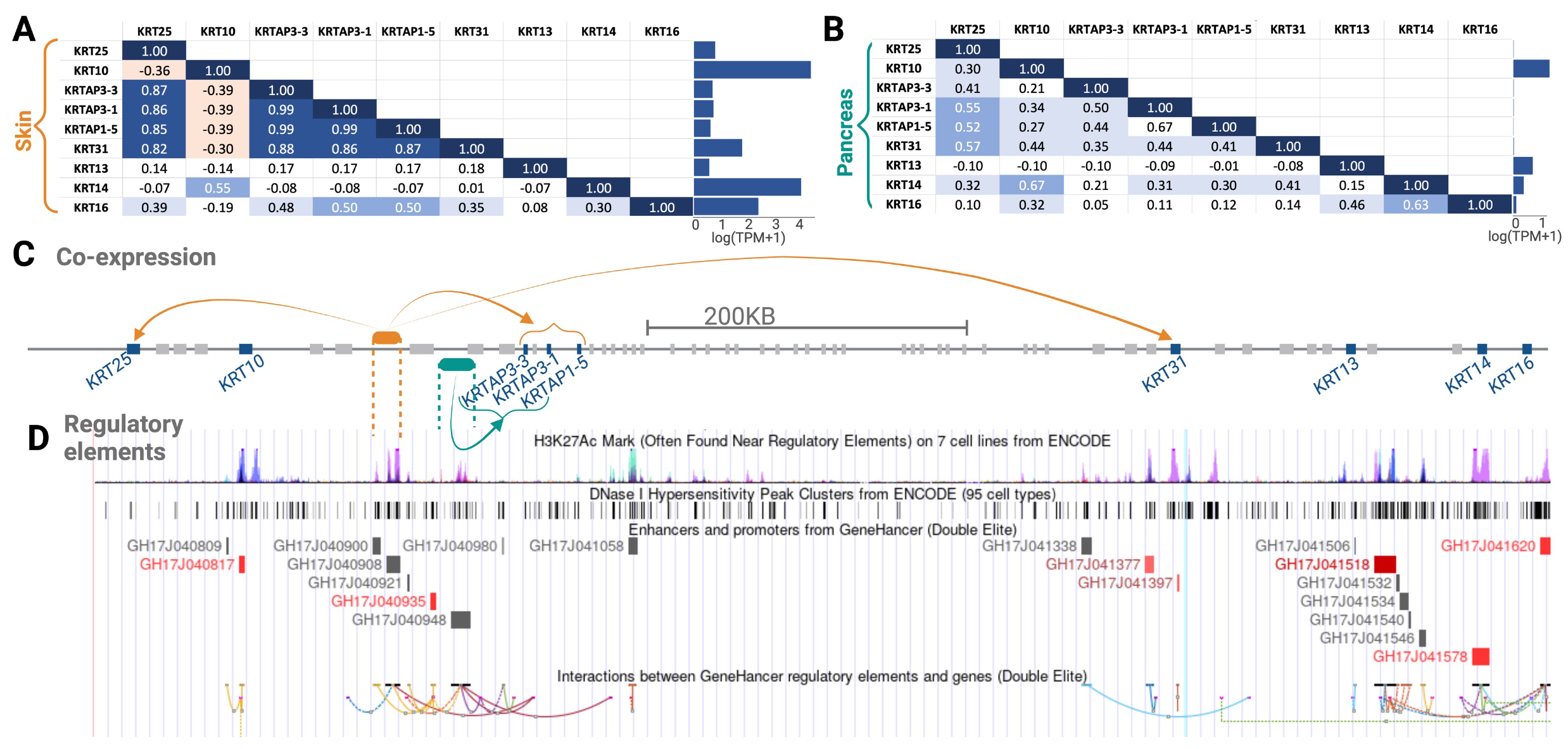}
\caption{{\bf Expression and co-expression analysis of a cluster of genes in community 5 on chromosome 17.} The co-expression matrices for these genes in (A) skin and in (B) pancreas are shown. The average expression for each gene in these tissues is shown in the bar graphs. The location of these genes on chromosome 17 is shown in (C), with arrows (colored according to the associated tissue) pointing from putative regulatory elements to highly co-expressed genes. (D) The panel shows different measures of the regulatory potential of this genome section. From top to bottom: 1. H3K27AC modification to histone H3 within the region, which often correlates with activation of transcription and is associated with active enhancers in a given tissue available through ENCODE database \cite{encode2012integrated}.  2. DNAse1 hypersensitivity sites. They are sections of the genome that are cut by DNAse1 enzyme. Given that the chromatin has to be ``open'' for the DNAse to access the sequence, the sequences that are cut by DNAse indicate open chromatin, which is in turn associated with regulatory activity. Data are available through ENCODE database \cite{encode2012integrated}. 3. Enhancer/promoters. These are sequences that are predicted as enhancers (gray) and promoters (red) from the GeneHancer database \cite{fishilevich2017genehancer}. 4. Established interactions between regulatory regions and genes as documented by GeneHancer database \cite{fishilevich2017genehancer}. These data sets combined with our co-expression analysis provide a novel outlook into potential topologically associated domains that may be regulated by specific sequences in a tissue-specific manner.
}
\label{fig:KRT}
\end{figure}

After initial investigation of the three gene clusters in community 5, i.e., one each on chromosome 4, 11, and 17, we further analyzed the one on chromosome 17, because keratin loci have been discussed in the context of human evolution \cite{eaaswarkhanth2014geographic, ho2022update}.
We show in Fig~\ref{fig:KRT}A and~\ref{fig:KRT}B the expression of each gene in this gene cluster in the skin and pancreas, respectively. We found that gene expression trends vary between the two tissues. Specifically, our method identified community 5 because of co-expression trends in the pancreas. However, in terms of the sheer expression level, the present gene cluster is expressed multiple folds higher in the skin than pancreas. Further, we found that the co-expression patterns for some gene pairs within this gene cluster are common between the skin and pancreas but differ for other gene pairs. The physical clustering of the genes that are co-expressed implicates genetic variation in shared gene regulatory factors as the main basis for co-expression. For example, a search of the GTEx eQTL database showed that the common single nucleotide polymorphism rs12450846 is significantly ($p<10^{-18}$) associated with lower expression of \textit{KRT31} in the skin but higher expression of this gene in ovaries ($p<0.005$). Unfortunately, this analysis was not conducted in the pancreas. Regardless, this polymorphism and the haplotype linked to it regulate multiple other keratins and keratin-associated protein genes in this particular locus in a tissue and gene-specific manner according to the GTEx database. Thus, genetic variation that affects the efficacy of regulatory regions (Fig~\ref{fig:KRT}C) or the formation of topologically associated domains (Fig~\ref{fig:KRT}D) in a tissue-specific manner may underly the co-expression of the genes in community 5 on chromosome 17. Indeed, we found several topologically associated domains, enhancers, transcription factor binding sites, and open chromatins within this region, affecting co-expressed genes in a similar fashion (Fig~\ref{fig:KRT}). Overall, our analysis provides several exciting hypotheses for future work to investigate regulatory regions that target multiple nearby genes and explain tissue-specific co-expression trends.

Another interesting community we identified is community 7. The genes in this community are located on different chromosomes and are enriched for response to temperature change (adaptive thermogenesis; see Table E in \nameref{S1Text}). Because they exist on different chromosomes, it is unlikely that these genes share any common regulatory sequences or topologically associating domains. Instead, their co-expression may be due to environmental stimuli that are shared among the samples at the time of sampling (e.g., warm or cold environments). If true, the co-expression is due to a response to environmental stimuli that is controlled by specific regulatory sequences with broad effects across the genome, such as transcription factors. Thus, our network analysis may be useful for identifying gene clusters that respond to different environments. 

\subsection{Gene pairs with shared associated SNPs in pancreas} \label{eqtl_results}

As described in section~\ref{eqtl_methods}, we hypothesized that genetic variation that affects gene expression in a tissue-specific manner can explain some of the co-expression trends we observed. Identifying such variation is challenging because of the huge amount of combinations that are possible between genetic variants and gene expression levels. To overcome this challenge and identify examples of where genetic variation may explain the co-expression trends and chromosomal clustering, we conduct an eQTL  analysis considering only cis variants that are physically close to genes of interest. This analysis provides a list of variants (SNPs in this case) that are significantly associated with expression levels of nearby genes. We will refer to these SNPs as eQTLs. Using this approach, we identified three gene pairs (i.e., \textit{CELA3B} and \textit{CELA3A}; \textit{AMY2B} and \textit{AMY2A}; \textit{REG3G} and \textit{REG1B}) that share associated eQTLs in the pancreas out of all the gene pairs in the network of $203$ genes with co-expression greater than $0.5$.

Notably, out of these three gene pairs, two pairs, i.e., the \textit{CELA3B}-\textit{CELA3A} and \textit{AMY2B}-\textit{AMY2A} pairs, are not composed of hub genes within the pancreas single-layered network and are only identified through our multilayer network approach. Both pairs are within community 5. For example, if we searched for the top 86 genes in terms of the weighted degree in the pancreas to match the number of genes in community 5, we were not able to identify the \textit{CELA3B}-\textit{CELA3A} or \textit{AMY2B}-\textit{AMY2A} pairs.  In contrast, the other gene pair with shared eQTLs (i.e., \textit{REG3G} and \textit{REG1B}) consists of two hub genes in the single-layered pancreas co-expression network. Therefore, we would have missed two out of three gene pairs that may be biologically interesting if we simply investigated hub genes in the pancreas.

\begin{figure}[!h]
\centering
\includegraphics[width=\linewidth]{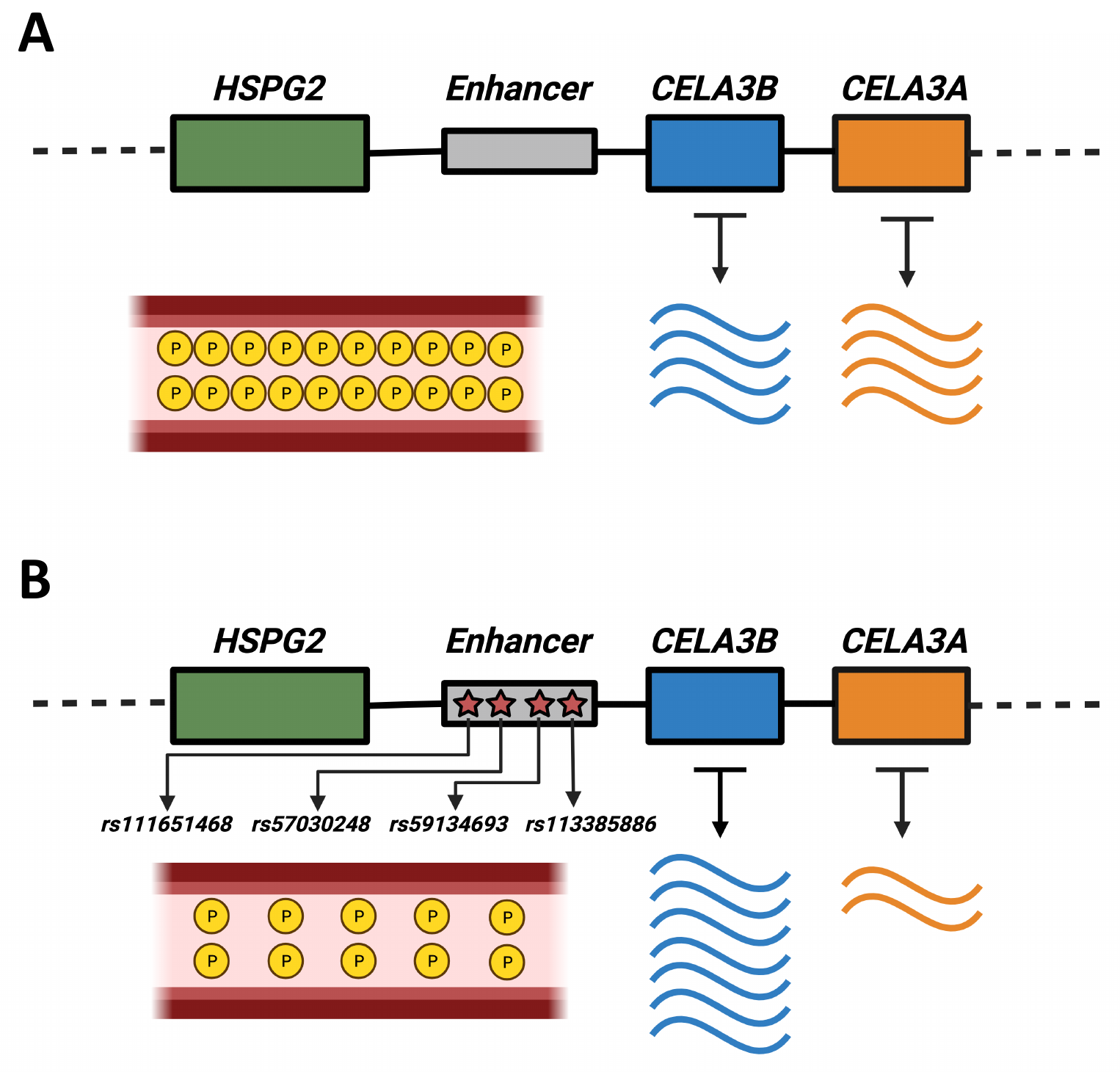}
\caption{{\bf A schematic of SNPs in an enhancer region (gray box) that affect the expression of \textit{CELA3A} (blue box) and \textit{CELA3B} (orange box) in the pancreas and are associated with blood phosphate concentration.}  (A) Expression levels of \textit{CELA3A} and \textit{CELA3B}, and blood phosphate concentration when the derived alleles for the putatively causal SNPs are absent. (B) The presence of the derived alleles for the putatively causal SNPs decreases the expression level of \textit{CELA3A}, increases the expression level of \textit{CELA3B}, and decreases the blood phosphate concentration.}
\label{fig:enhancer}
\end{figure}

Next, to identify the biological relevance of this putatively genetically determined co-expression pattern, we investigated the CELA3 locus. We identified a set of 96 variants from statistically significant eQTLs for both \textit{CELA3A} and \textit{CELA3B} in the pancreas. CELA3A and CELA3B, which are proteases, are produced as zymogens in the pancreas. They then perform their digestive function in the intestine once they have been transported there. It has previously been speculated that the presence of two \textit{CELA3} copies provides a functional substitute for the lack of pancreatic expression of CELA1 in humans relative to pigs \cite{boros2017overlapping}. The 96 variants are present in the genomic region spanned by \textit{HSPG2}, \textit{CELA3A}, and \textit{CELA3B}. The minor allele for each of these 96 variants is associated with a decreased expression of \textit{CELA3A} and an increased expression of \textit{CELA3B} in the pancreas. This observation may hint at a possible constraint on the combined expression level of \textit{CELA3A} and \textit{CELA3B} in the pancreas, further supporting the idea that CELA genes may have compensatory roles for the functions of other members in this gene family. To understand the population genetics trends affecting the regulatory variants that we identified, we analyzed 83 SNPs that are associated with gene expression of \textit{CELA3A} and \textit{CELA3B} and genotyped in the 1000 Genomes Project Phase-3 data set. We found that these variants form a single linkage-disequilibrium (LD) group in Europeans at an $r^2$ threshold of 0.6 \cite{takeuchi2005linkage}. The minor alleles of 10 of these variants are associated with a decreased blood phosphate concentration \cite{biobank, aqil2023balancing} (see Fig~\ref{fig:enhancer}). In order to identify putative causal variants in the LD group, we investigated whether any of these variants lie in a regulatory region. We find that four (rs57030248, rs59134693, rs113385886, and rs111651468) of these variants lie in an enhancer (ENSR00000350171), identified by ENSEMBL's variant effect predictor \cite{mclaren2016ensembl} (Fig~\ref{fig:enhancer}B), which is active in the pancreas. Three of these four variants (rs57030248, rs59134693, and rs113385886) are both present in the enhancer region and associated with decreased blood phosphate levels. It is likely that one or more of these three variants are causal in the context of differences in the expression levels of \textit{CELA3A} ($p=1.1\cdot 10^{-8}$, normalized effect size $=-0.43$) and \textit{CELA3B} ($p=4.5\cdot 10^{-10}$, normalized effect size $=+0.43$). Our results allowed us to construct a hypothetical model (Fig~\ref{fig:enhancer}). Our multilayer network approach facilitated the narrowing down of putatively causal genetic variants that affect the expression levels of negatively co-expressed gene pairs within the context of protein and phosphate metabolism.

\section{Discussion} \label{discussion}
We developed a multilayer community detection method for Pearson correlation matrix data. We applied the proposed method to gene co-expression data from four tissues in humans to identify gene modules (i.e., communities). Some detected communities spanned multiple layers, which we refer to as generalist communities. Other communities lay mostly within one layer, specifically the pancreas layer, which we refer to as specialist communities. We then found that both generalist and specialist communities were localized on a smaller number of chromosomes than the expectation of random distribution of genes. As a case study, we closely looked into two groups of genes (i.e., the KRTAP cluster in community 5 and community 7 as a whole) and suggested that the detected multilayer communities may imply gene regulatory factors shared across different tissues or environmental stimuli shared among samples. Finally, we found three gene pairs that share associated eQTLs in the pancreas, identifying examples in which genetic variation may explain the co-expression trends and chromosomal clustering.

Various mutually inclusive factors can explain co-expression of genes \cite{leek2007capturing, gaiteri2014beyond, van2018gene}.
We explored two such factors in our case study. First, it is possible that the regulatory regions control the expression of multiple genes in certain tissues \cite{allocco2004quantifying, hurst2004evolutionary, sproul2005role, ebisuya2008ripples, perry2022snake}. 
In this case, individuals who share genetic variations in these regulatory regions will have similar expression levels in these tissues where these regulators are active. If genetic variation underlies the co-expression of genes and the regulatory elements are cis (i.e., close physical proximity), we expect the co-expressed genes to cluster across the genome. We suggested that \textit{KRTAP3-3}, \textit{KRTAP3-1}, and \textit{KRTAP1-5} share regulatory elements in skin and pancreas. Indeed, several recent studies highlight topologically associating domains as potential sites underlying co-expression of multiple proximate genes \cite{saitou2020functional, perry2022snake}. Our approach integrated with chromatin accessibility (e.g., ATAC-seq) data is expected to facilitate identifying such loci where regulatory architecture may underlie the gene expression trends of multiple nearby genes in a tissue-specific manner. 
Second, it is possible that co-expressed genes have similar or complementing functions that respond to particular environmental conditions \cite{holter2000fundamental, carter2004gene}.
For example, we suggested that the genes in community 7 detected in the multilayer correlation matrix with $\gamma=3$ may be involved in response to temperature change and co-expressed because samples were subjected to respective environmental conditions at the time of sampling. 
We argue that the response to environmental stimuli may underlie co-expression in these genes and thus indicate phenotypic plasticity for related traits \cite{fusco2010phenotypic}, where an individual can respond to different environmental cues by adjusting the expression levels of multiple genes \cite{gibert2016phenotypic}.
 Our approach can provide a systematic framework to study phenotypic plasticity using animal models comparing different environmental stimuli (e.g., temperature, pathogenic pressure, diet, xenobiotic substances).

Another particularly relevant study using GTEx data to construct tissue-specific gene co-expression networks compared the community structure across different tissues \cite{azevedo2021multilayer}. While the present study also uses GTEx data to construct tissue-specific gene co-expression networks and compare community structure across layers, the details of the methods differ in the following noteworthy ways. Azevedo et al.\,\cite{azevedo2021multilayer} apply a thresholding method to the correlation matrices to construct networks and use signed modularity \cite{gomez2009analysis} as the quality function for community detection, whereas the present study uses the correlation matrices directly with an appropriate correlation matrix null model in the quality function, as described in section \ref{commdetectioncorr}. Additionally, we perform a multilayer community detection method that incorporates interlayer coupling strength information, whereas Azevedo et al.\,perform single layer community detection on each layer separately and then compare the community structure across networks using the global multiplexity index \cite{hristova2016international}. The global multiplexity index quantifies how many times two genes belong to the same communities across all the layers. To connect terminology in their study \cite{azevedo2021multilayer} and the present study, we point out that a group of genes with global multiplexity index equal to $\mathcal{L}$ (i.e., the total number of layers) corresponds to a generalist community that spans all layers of the multilayer network. This type of community is also called a pillar community~\cite{hanteer2020unspoken}. A group of genes with global multiplexity index equal to $1$ corresponds to a specialist community. Finally, a group of genes with global multiplexity index greater than $1$ but less than $\mathcal{L}$ is a generalist community that spans a subset of the layers (of size equal to the global multiplexity index) in the multilayer network. This type of community is also called a semi-pillar community~\cite{hanteer2020unspoken}. Both Azevedo et al.\,\cite{azevedo2021multilayer} and the present study employ enrichment analysis on the communities to identify known biological processes corresponding to the discovered gene communities. Systematic comparison between multilayer community detection methods, such as the present work, and single layer community detection methods with multilayer analysis, such as \cite{azevedo2021multilayer}, warrants future work.

We employed multilayer modularity maximization. By design, modularity maximization consists of finding an optimal partition of nodes into non-overlapping communities, and therefore each node belongs to exactly one community. This feature is inherited to multilayer modularity maximization such that each node $(i, \alpha)$, where $i$ represents a gene and $\alpha$ represents a layer, belongs to exactly one community. Multilayer modularity maximization has been used on biological networks to extract groups of proteins or genes that may be functionally related. For example, this technique was used on multilayer networks composed of transcription factor co-targeting, microRNA co-targeting, PPI, and gene co-expression networks as four layers for revealing candidate driver cancer genes \cite{cantini2015detection}
and on a multilayer network composed of pathways, co-expression, PPIs, and complexes networks for obtaining groups of disease-related proteins \cite{didier2015identifying}.
However, it is not straightforward to interpret the obtained multilayer communities as gene module because, within a single multilayer community, different genes appear in different sets of layers. For example, in a generalist community spanning all the four layers, some genes $i$ may be present in all the layers, whereas other genes $j$ may be present in only one layer. Then, although $i$ and $j$ belong to the same community and connected by group-level co-expression relationships, it may be difficult to argue that $i$ and $j$ share biological functions or environmental factors because how their co-expression depends on layers is different between genes $i$ and $j$. 
One option to mitigate this problem is to focus on the resulting gene set in a given multilayer community and ignore the layer identity for simplicity  \cite{cantini2015detection, didier2015identifying}. In contrast, we limited our analysis of generalist communities to the genes that appear in at least three out of the four layers in the community. In this manner, we argued that the genes in the generalist communities used in our localization and biological analyses may have functions common across different tissue types. For the two specialist communities that we analyzed in depth (with $\gamma=3$), we did not need to select genes because all genes were present in the pancreas and only a small fraction of genes were also present in a different tissue type. 

The GTEx Consortium portal provides gene expression data from 30 types of tissues \cite{lonsdale2013genotype}. It is computationally straightforward to extend this analysis to more than four layers (i.e., tissues). Then, however, the results would quickly become much more complicated to interpret. With a number of layers much larger than four, it is likely that our method would no longer discover specialist communities.
This is an important limitation of the present analysis. 
Developing methods more directly tailored to multilayer gene co-expression networks and correlation matrices with a larger number of tissues warrants future work. A suitable method should depend on biological questions. For example, enforcing pillar or semi-pillar communities such that all the genes belonging to the same multilayer community are present in the same set of layers \cite{hanteer2020unspoken, magnani2021community} may facilitate biological interpretation of obtained results. Allowing overlapping of communities \cite{tripathi2019adapting, riccio2021identifying}
and genes not belonging to any community may be another choice. For example, overlapping community detection in single-layer networks has been shown to be better at identifying biologically relevant disease modules than non-overlapping community detection~\cite{tripathi2019adapting}.

We only analyzed co-expression among $N=203$ out of the $56,200$ genes because it is difficult to reliably estimate covariance matrices when the number of samples is small \cite{dempster1972covariance, johnstone2001distribution, pierson2015sharing, ballouz2015guidance, lyu2018condition}. 
Justifiable methods for analyzing co-expression matrices or networks of a larger number of genes are desirable. Such methods will enable us to reduce bias involved in choosing a small subset of genes to analyze. In contrast, a different approach is to formulate the estimation of large correlation networks from big data as a computational challenge and work on efficient algorithms and application to complex biological data \cite{becker2023large}. Systematically investigating biological performance of network community detection as a function of the number of samples \cite{ballouz2015guidance, guo2017evaluation, ovens2020impact} will help us to better understand potentials and limitations of both single-layer and multilayer community detection in gene and other related networks, which is left as future work.

\section*{Data Availability}
All data that are used in the study can be found publicly. The references and databases are provided in the manuscript. Direct link to GTEx TPM data: \url{https://storage.cloud.google.com/adult-gtex/bulk-gex/v8/rna-seq/GTEx_Analysis_2017-06-05_v8_RNASeQCv1.1.9_gene_tpm.gct.gz}. Direct link to GTEx eQTL data: \url{https://storage.cloud.google.com/adult-gtex/bulk-qtl/v8/single-tissue-cis-qtl/GTEx_Analysis_v8_eQTL.tar}. Code is available in a GitHub repository: \url{https://github.com/russell-madison/corr_comm_detection}.

\section*{Supporting information}
\paragraph*{S1 Text.}
 \label{S1Text}
{\bf Fig A. Composition of each community by layer, i.e., tissue, for the multilayer correlation matrix originating from the $150$ genes with the highest variance of TPM in each tissue, detected with our community detection method for multilayer correlation matrices with $\gamma=3$.} Although there are $50$ communities detected, we only show the communities with more than one gene in this figure. The darker shades indicate nodes corresponding to genes that only appear in one layer in the given community. The lighter shades indicate genes corresponding to genes that appear in multiple layers in the community. 
{\bf Fig B. Jaccard index between the set of tissue-specific hub genes and the set of genes in a community.} Each row corresponds to the top 50 hub genes in each layer (i.e., tissue), where ``panc'' denotes pancreas, ``sal'' denotes salivary gland, ``mamm'' denotes mammary gland, and ``skin'' denotes skin (not sun exposed). Each column corresponds to a community identified with $\gamma=3$.
{\bf Table A. Z scores for the number of intralayer edges within each community and for the conductance of each community detected in the unweighted multilayer network obtained by graphical lasso with $\gamma=1$ and $\gamma=3$.} Comm.\,denotes community and no.\,denotes "number of".
{\bf Table B. Z scores for the average distance between pairs of genes on each chromosome and each significant community detected with $\gamma=1$.} Comm.\,denotes community and Chr denotes chromosome.
{\bf Table C. Z scores for the average distance between pairs of genes on each chromosome and each significant community detected with $\gamma=3$.} Comm.\,denotes community and Chr denotes chromosome.
{\bf Table D. Results of the gene set enrichment analysis for the top 50 highly expressed genes out of the $203$ genes in the network in each tissue.}
{\bf Table E. Results of the gene set enrichment analysis for the communities of the multilayer correlation matrix with $\gamma=3$.} Comm.\,denotes community.
{\bf Table F. Results of the gene set enrichment analysis for the top 50 highly connected genes out of the $203$ genes in the single-layer network of each tissue.}
{\bf Text A. Analysis of an expanded multilayer correlation matrix.}
{\bf Text B. Significance of communities detected in general multilayer networks.}
{\bf Text C. Derivation of the variance of the total intralayer weight for a community in a multilayer correlation matrix.}
{\bf Text D. Graphical lasso.}
{\bf Text E. Z scores for the average distance between pairs of genes on each chromosome separately in each community.}
{\bf Text F. Results of the gene set enrichment analysis.}
{\bf Text G. Tissue-specific hub genes versus gene communities.}

\section*{Author Contributions}
\textbf{Conceptualization:} Marie Saitou, Omer Gokcumen, Naoki Masuda. \newline

\noindent \textbf{Data curation:} Madison Russell, Alber Aqil, Marie Saitou. \newline

\noindent \textbf{Formal analysis:} Madison Russell, Alber Aqil, Omer Gokcumen. \newline

\noindent \textbf{Funding acquisition:} Omer Gokcumen, Naoki Masuda. \newline

\noindent \textbf{Investigation:} Madison Russell, Alber Aqil, Marie Saitou, Omer Gokcumen, Naoki Masuda. \newline

\noindent \textbf{Methodology:} Madison Russell, Alber Aqil, Naoki Masuda. \newline

\noindent \textbf{Project administration:} Naoki Masuda. \newline

\noindent \textbf{Resources:} Marie Saitou, Omer Gokcumen. \newline

\noindent \textbf{Software:} Madison Russell. \newline

\noindent \textbf{Supervision:} Naoki Masuda. \newline

\noindent \textbf{Validation:} Madison Russell. \newline

\noindent \textbf{Visualization:} Madison Russell, Alber Aqil, Omer Gokcumen. \newline

\noindent \textbf{Writing \textendash{} original draft:} Madison Russell, Alber Aqil, Omer Gokcumen, Naoki Masuda. \newline

\noindent \textbf{Writing \textendash{} review \& editing:} Madison Russell, Alber Aqil, Marie Saitou, Omer Gokcumen, Naoki Masuda.



\clearpage

\begin{center}
\vspace*{12pt}
{\Large Supplementary Information for:\\
\vspace{12pt}
Gene communities in co-expression networks across different tissues}
\vspace{12pt} \\
\end{center}

\setcounter{figure}{0}
\setcounter{table}{0}
\setcounter{section}{0}
\setcounter{equation}{0}

\renewcommand{\thesection}{S\arabic{section}}
\renewcommand{\thefigure}{S\arabic{figure}}
\renewcommand{\thetable}{S\arabic{table}}
\renewcommand{\theequation}{S\arabic{equation}}

\begin{center}
\author{Madison Russell, Alber Aqil, Marie Saitou, Omer Gokcumen, Naoki Masuda}
\vspace{24pt} \\
\end{center}

\section*{S1 Text} \label{SI}

\subsection*{Text A. Analysis of an expanded multilayer correlation matrix}\label{expanded_network}

To validate our choice of the top $75$ genes in each tissue in terms of the variance of TPM, we repeated the same analysis with the top $150$ genes in each tissue in terms of the variance of TPM. The union of the top $150$ genes across the four tissues contains $371$ genes. We analyze a four-layer correlation matrix composed of these $371$ genes. We run our community detection method for multilayer correlation matrices with $\gamma=3$, which is the main value of $\gamma$ used in the analysis in the main text. We show the partition of the $371$-gene multilayer correlation matrix in Fig \ref{fig:expanded_partition}.

\renewcommand{\thefigure}{A}
\captionsetup{font={small,rm}} 
\captionsetup{labelfont=bf}
\begin{figure}[H]
  \begin{center}
    \includegraphics[width=\textwidth]{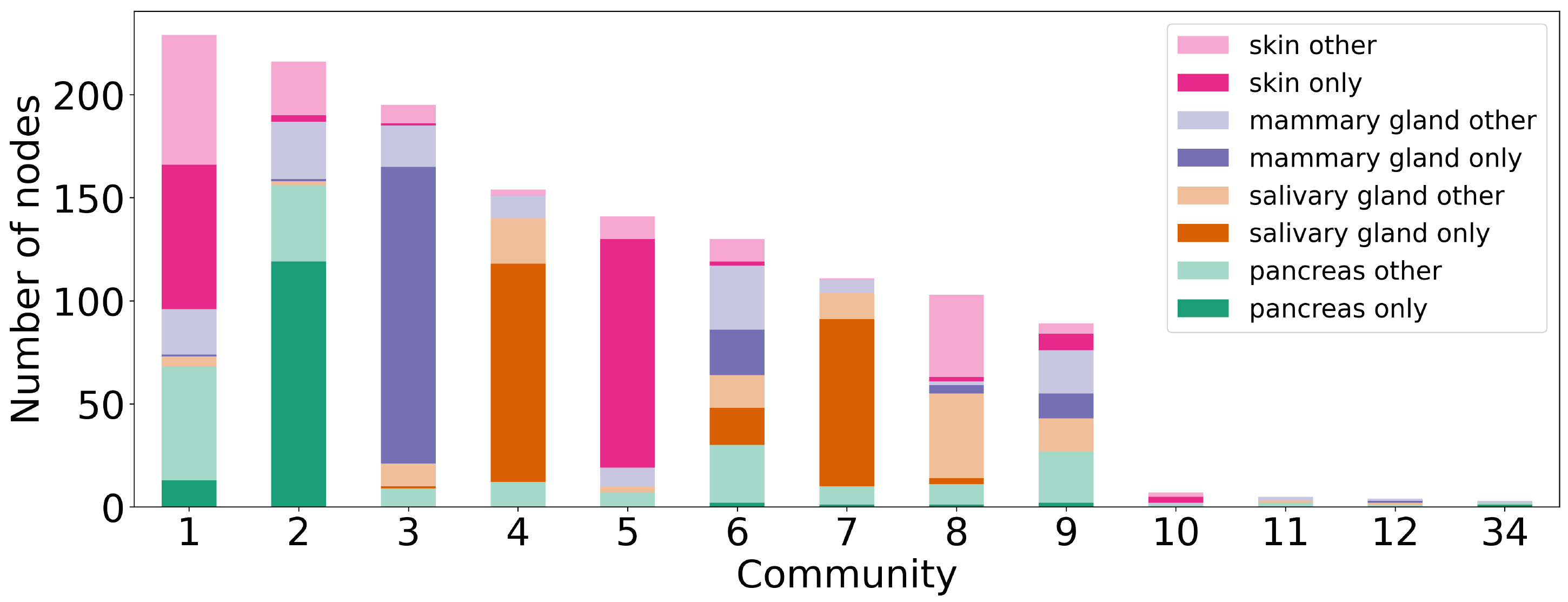}
  \end{center}
  \caption{Composition of each community by layer, i.e., tissue, for the multilayer correlation matrix originating from the $150$ genes with the highest variance of TPM in each tissue, detected with our community detection method for multilayer correlation matrices with $\gamma=3$. Although there are $50$ communities detected, we only show the communities with more than one gene in this figure. The darker shades indicate nodes corresponding to genes that only appear in one layer in the given community. The lighter shades indicate genes corresponding to genes that appear in multiple layers in the community.}
  \label{fig:expanded_partition}
\end{figure}

We find that the expanded multilayer correlation matrix with $371$ genes produces a similar type of partition as the original multilayer correlation matrix with $203$ genes, in the sense that there are some specialist communities and some generalist communities. We compare the larger of the pancreas specialist communities in the original correlation matrix (i.e., community 5 in the main text, shown in Fig 3(b)) to the pancreas specialist community in the expanded correlation matrix (i.e., community 2 in Fig \ref{fig:expanded_partition}). Community 5 in the original correlation matrix contains $86$ genes in the pancreas layer, and $88\%$ of these genes also appear in community 2 in the pancreas layer of the expanded correlation matrix. Hence, there is significant overlap between the pancreas specialist communities of the two correlation matrices. This result supports robustness of our analysis with respect to the choice of the number of genes selected for our analysis.

\subsection*{Text B. Significance of communities detected in general multilayer networks} \label{SI1}

\renewcommand{\thesubsubsection}{S3.1}
\subsubsection*{B.1 Null model} \label{SI1null}

We assume undirected and unweighted networks. A common choice for a null model is the configuration model, in which the degree sequence $\{k_i\}_{i=1}^{N}$ is specified, and an edge is laid between nodes $i$ and $j$ with probability 
\begin{equation} \label{eq:pij}
p_{ij}=\frac{k_i k_j}{2M},\tag{S1}
\end{equation}
where $k_i$ is the degree of the $i$th node in the original network. However, we avoid the configuration model for two reasons.
First, by imposing the value of $k_1, \ldots, k_N$, the stochastic generation of edges sharing a node is not independent of each other. For example,
if $k_1$ is small and we have generated edge $(1, 2)$ with probability $p_{12}$, then edge $(1, 3)$ is generated with a probability smaller than $p_{13}$.
This type of correlation makes it difficult to analytically derive the quality measure of individual communities that requires the count the edges sharing a node.
Second, large $k_i k_j$ values can yield $p_{ij}>1$ \cite{park2003origin}. One could set a structural cutoff degree $k_{\text{max}}$ to be of the order of $\sqrt{N}$ to enforce the constraint $p_{ij}<1$ \cite{catanzaro2005generation}. However, it is often the case that the largest degree in an empirical network far exceeds this structural cutoff value~\cite{maslov2004detection, garlaschelli2009generalized}.

Instead of the configuration model, we use an exponential random graph model (ERGM) as the null model. Instead of the exact degree, we fix the expected degree of each $i$th node to $k_i^*$, the degree of the same node in the original network. Let $\Omega$ be an ensemble of networks with $N$ nodes. Let $\vec{\theta} \equiv (\theta_1, \ldots, \theta_N)$ be the model parameters. The probability distribution of the adjacency matrix, $A=(A_{ij})$, that maximizes the Shannon entropy subject to the constraints 
\begin{equation} \label{eq:constraints}
\sum_{A\in\Omega}P(A)k_{i}(A)=k_i^*,\tag{S2}
\end{equation}
where $k_{i}(A)$ is the degree of the $i$th node in network $A$, and the normalization condition
\begin{equation} \label{eq:norm}
\sum_{A\in\Omega}P(A)=1\tag{S3}
\end{equation}
is
\begin{equation} \label{eq:graphprob}
P(A|\vec{\theta})=\prod_{i=1}^{N}\prod_{j=1}^{i-1} p_{ij}^{A_{ij}}(1-p_{ij})^{(1-A_{ij})},\tag{S4}
\end{equation}
where
\begin{equation} \label{eq:probij}
p_{ij}=\frac{e^{-\theta_{i}-\theta_{j}}}{1+e^{-\theta_{i}-\theta_{j}}}\tag{S5}
\end{equation}
is the probability that nodes $i$ and $j$ are adjacent~\cite{park2004statistical, cimini2019statistical, vallarano2021fast}.

We infer the model parameters $\vec{\theta}$ by maximizing the associated log-likelihood function, i.e., 
\begin{equation} \label{eq:loglikelihood}
\mathscr{L}(\vec{\theta})\equiv \ln P(A^*|\vec{\theta})=-\sum_{i=1}^{N}\theta_{i}k_{i}^*-\sum_{i=1}^{N}\sum_{j=1}^{i-1}\ln(1+e^{-\theta_{i}-\theta_{j}}),\tag{S6}
\end{equation}
where $A^*$ is the adjacency matrix of the original network. One can derive Eq.~\eqref{eq:loglikelihood} from Eq.~\eqref{eq:graphprob}.

We numerically determine $\theta_{i}$'s using a fixed point method implemented in the Python package NEMtropy \cite{vallarano2021fast}. We use NEMtropy to generate this so-called undirected binary configuration model (UBCM) \cite{vallarano2021fast}. We estimate the UBCM for each layer independently. Using the obtained multilayer UBCM as null model, we calculate the statistical significance of individual communities detected in the original multilayer network.

\subsubsection*{B.2 Number of intralayer edges within each community} \label{SI1intralayer}

We use two measures to assess the quality of individual communities. The first measure is the number of intralayer edges within each community, which is essentially the same as $X$ used in the main text for correlation matrices. Previous studies used the number of edges within a community in single-layer networks \cite{radicchi2004defining, yang2015defining}, and we extend this quality measure to multilayer networks. 

Let $S$ be a set of nodes in a multilayer network. Let $X_{\alpha}$ be the number of edges within $S$ in layer $\alpha$, and let $X$ be the total number of intralayer edges within $S$, i.e.,
\begin{equation} \label{eq:X}
X=\sum_{\alpha=1}^{\mathcal{L}}X_{\alpha}.\tag{S7}
\end{equation}
In the UBCM, the edges are independently laid. Therefore, $X$ obeys the Poisson binomial distribution, which is the discrete probability distribution of a sum of independent Bernoulli trials that are not necessarily identically distributed \cite{wang1993number}.

Let $\theta_{i\alpha}$ be the UBCM parameter for node $i$ in layer $\alpha$. Using Eq.~\eqref{eq:probij} for each layer, we obtain the expectation of $X$ as follows:
\begin{equation} \label{eq:expedges}
  \text{E}[X]=\sum_{\alpha=1}^{\mathcal{L}}\sum_{\substack{i=1 \\ (i,\alpha)\in S}}^{N}\sum_{\substack{j=1 \\ (j,\alpha)\in S}}^{i-1} \frac{e^{-\theta_{i\alpha}-\theta_{j\alpha}}}{1+e^{-\theta_{i\alpha}-\theta_{j\alpha}}},\tag{S8}
\end{equation}
where the summation is over all node pairs $(i,\alpha),(j,\alpha)$ in $S$. Note that we have excluded self-loops.
The variance of $X$ is equal to~\cite{wang1993number} 
\begin{equation} \label{eq:varedges}
 \text{Var}[X]=\sum_{\alpha=1}^{\mathcal{L}}\sum_{\substack{i=1 \\ (i,\alpha)\in S}}^{N}\sum_{\substack{j=1 \\ (j,\alpha)\in S}}^{i-1}\left(1-\frac{e^{-\theta_{i\alpha}-\theta_{j\alpha}}}{1+e^{-\theta_{i\alpha}-\theta_{j\alpha}}}\right) \frac{e^{-\theta_{i\alpha}-\theta_{j\alpha}}}{1+e^{-\theta_{i\alpha}-\theta_{j\alpha}}}.\tag{S9}
\end{equation}

\subsubsection*{B.3 Conductance of each community} \label{SI1cond}

The second quality measure is the conductance of each community. Let $G(V,E)$ be an undirected single-layer network, and we consider a set of nodes $S\subseteq V$. Let
\begin{equation}
c_{s} = \left| \{(u,v)\in E: u\in S, v\notin S\} \right|\tag{S10}
\end{equation}
be the number of edges on the boundary of $S$ and 
\begin{equation}
m_{s} = \left| \{(u,v)\in E: u\in S, v\in S\} \right|\tag{S11}
\end{equation}
be the number of edges within $S$. Then, the conductance of $S$ is given by \cite{yang2015defining}
\begin{equation} \label{eq:singcond}
\varphi(S)=\frac{c_{s}}{2m_{s}+c_{s}}.\tag{S12}
\end{equation}
The conductance measures the fraction of the number of half-edges emanating from nodes in $S$ that are connected to a half-edge emanating from a node outside $S$. Therefore, the conductance is small for a good community \cite{shi2000normalized}.

To define the conductance of a set of nodes $S$ in a multilayer network with $\mathcal{L}$ layers, let $Y_{\alpha}$ be the number of edges on the boundary of $S$ in layer $\alpha$. We define the conductance of $S$ by
\begin{equation} \label{eq:multcond}
\varphi(S)=\frac{\sum_{\alpha=1}^{\mathcal{L}}Y_{\alpha}}{\sum_{\alpha=1}^{\mathcal{L}}(2X_{\alpha}+Y_{\alpha})}=\frac{Y}{2X+Y},\tag{S13}
\end{equation}
where $Y$ is the number of intralayer edges on the boundary of $S$. Because the UBCM independently lays edges for different node pairs, $Y$ as well as $X$ obeys a Poisson binomial distribution. It should also be noted that $X$ and $Y$ are independent because they are calculated based on disjoint sets of node pairs.

We denote the set of intralayer node pairs within community $S$ by
\begin{equation}
E_{\max}^{\rm within} = \{ (i,j,\alpha) : (i, \alpha) \in S, (j, \alpha) \in S, i < j\}.\tag{S14}
\end{equation}
We note that the cardinality (i.e., number of elements) of $E_{\max}^{\rm within}$ is equal to
\begin{equation}
x_{\max} \equiv \sum_{\alpha=1}^{L} \frac{N'_{S\alpha}(N'_{S\alpha}-1)}{2},\tag{S15} 
\end{equation}
where $N'_{S\alpha}$ is the number of nodes in $S$ in layer $\alpha$.
Because $X$ obeys the Poisson binomial distribution, the probability for $X$ is
\begin{equation} \label{eq:pmfx}
P_{X}(X=x)= \sum_{\overline{E}\subset E_{\max}^{\rm within} \text{ s.t. } \left|\overline{E}\right| = x}
\prod_{(i,j,\alpha) \in \overline{E}} p_{ij\alpha} \prod_{(i,j,\alpha)\in E_{\max}^{\rm within} \setminus \overline{E}}(1-p_{ij\alpha}),\tag{S16}
\end{equation}
where 
\begin{equation}
p_{ij\alpha} = \frac{e^{-\theta_{i\alpha}-\theta_{j\alpha}}}{1+e^{-\theta_{i\alpha}-\theta_{j\alpha}}},\tag{S17}
\end{equation}
and $x$ is the number of edges in $S$. Similarly, the probability for $Y$ is 
\begin{equation} \label{eq:pmfy}
P_{Y}(Y=y)=\sum_{\overline{E}\subset E_{\max}^{\rm boundary} \text{ s.t. } \left|\overline{E}\right| = y}
\prod_{(i,j,\alpha) \in \overline{E}}p_{ij\alpha} \prod_{(i,j,\alpha)\in E_{\max}^{\rm boundary} \setminus \overline{E}}(1-p_{ij\alpha}),\tag{S18}
\end{equation}
where
\begin{equation}
E_{\max}^{\rm boundary} = \{ (i, j, \alpha) : (i, \alpha) \in S, (j, \alpha) \notin S\}.\tag{S19}
\end{equation}
Note that the cardinality of $E_{\max}^{\rm boundary}$ is 
\begin{equation}
y_{\max}=\sum_{\alpha=1}^{\mathcal{L}}N'_{S\alpha}(N-N'_{S\alpha}).\tag{S20}
\end{equation}

Because $X$ and $Y$ are mutually independent, the expected value of the conductance of a set of nodes $S$ in the multilayer network is given by
\begin{equation} \label{eq:expcond}
\text{E}\left[\frac{Y}{2X+Y}\right]=\sum_{x=0}^{x_{\max}}\sum_{y=0}^{y_{\max}}\frac{y}{2x+y}P_{X}(X=x) P_{Y}(Y=y).\tag{S21}
\end{equation}
The variance of the conductance of $S$ is given by
\begin{align} \label{eq:varcond}
\text{Var}\left[\frac{Y}{2X+Y}\right] =& \sum_{x=0}^{x_{\max}}\sum_{y=0}^{y_{\max}}\left(\frac{y}{2x+y}\right)^{2} P_{X}(X=x) P_{Y}(Y=y) \notag\\
& -\left[\sum_{x=0}^{x_{\max}}\sum_{y=0}^{y_{\max}}\frac{y}{2x+y} P_{X}(X=x) P_{Y}(Y=y)\right]^{2}.\tag{S22}
\end{align}
In Eqs.~\eqref{eq:expcond} and \eqref{eq:varcond}, we set $\frac{y}{2x+y}=1$ for $(x, y) = (0, 0)$.

\subsubsection*{B.4 Results} \label{SI1results}

We show in Table \ref{table:lasso_zscores} the Z scores for the number of intralayer edges within each community and for the conductance of each community detected in the unweighted multilayer network obtained by graphical lasso with $\gamma=1$ and $\gamma=3$. For $\gamma=1$, communities 10 and 11 are each composed of a single gene; for $\gamma=3$, communities 11 through 14 are each composed of a single gene. We omitted these single-gene communities in Table \ref{table:lasso_zscores}. Except these single-gene communities, for both $\gamma=1$ and $\gamma=3$, all other communities are statistically significant with a large positive Z score for the number of intralayer edges
and a large negative Z score for the conductance.

\begin{table}[t]
\renewcommand\thetable{A}
\begin{center}
 \caption{Z scores for the number of intralayer edges within each community and for the conductance of each community detected in the unweighted multilayer network obtained by graphical lasso with $\gamma=1$ and $\gamma=3$. Comm.\,denotes community and no.\,denotes "number of".}
\begin{tabular}[b]{|c c c|c c c|}\hline
\multicolumn{3}{|c|}{$\gamma=1$} & \multicolumn{3}{c|}{$\gamma=3$} \\\hline
Comm. & \thead{Z score for no.\, \\ intralayer edges} & \thead{Z score for the \\ conductance} & Comm. & \thead{Z score for no.\, \\ intralayer edges} & \thead{Z score for the \\ conductance}\\ \hline
  1 & $26.787$ & $-35.697$ & 1 & $25.702$ & $-31.396$\\
  2 & $43.706$ & $-49.785$ & 2 & $50.807$ & $-60.543$\\
  3 & $41.245$ & $-49.208$ & 3 & $39.556$ & $-42.975$\\ 
  4 & $49.956$ & $-60.620$ & 4 & $32.401$ & $-34.112$\\
  5 & $42.446$ & $-45.851$ & 5 & $49.755$ & $-52.415$\\
  6 & $43.054$ & $-49.280$ & 6 & $48.585$ & $-52.584$\\
  7 & $34.795$ & $-37.814$ & 7 & $43.897$ & $-46.904$\\
  8 & $29.940$ & $-31.740$ & 8 & $37.580$ & $-40.672$\\
  9 & $57.664$ & $-62.551$ & 9 & $36.172$ & $-37.726$\\
   & &  & 10 & $30.597$ & $-32.125$\\ \hline
 \end{tabular}
 \label{table:lasso_zscores}
 \end{center}
 \end{table}

\subsection*{Text C. Derivation of the variance of the total intralayer weight for a community in a multilayer correlation matrix} \label{SI2}

To derive Eq.\,(11) in section 2.4, we start with
\begin{align} \label{eq:Vartotweight1}
& \text{Var}\left[\sum_{\alpha=1}^{\mathcal{L}}\sum_{\substack{i=1 \\ (i,\alpha)\in S}}^{N}\sum_{\substack{j=1 \\ (j,\alpha)\in S}}^{i-1}C_{ij\alpha}^{\text{con}}\right]\notag\\
=& \text{E}\left[\left(\sum_{\alpha=1}^{\mathcal{L}}\sum_{\substack{i=1 \\ (i,\alpha)\in S}}^{N}\sum_{\substack{j=1 \\ (j,\alpha)\in S}}^{i-1}C_{ij\alpha}^{\text{con}}\right)^2\right]-\left(\text{E}\left[\sum_{\alpha=1}^{\mathcal{L}}\sum_{\substack{i=1 \\ (i,\alpha)\in S}}^{N}\sum_{\substack{j=1 \\ (j,\alpha)\in S}}^{i-1}C_{ij\alpha}^{\text{con}}\right]\right)^2 \notag\\
{}=&\text{E}\left[\sum_{\alpha=1}^{\mathcal{L}}\sum_{\substack{i=1 \\ (i,\alpha)\in S}}^{N}\sum_{\substack{j=1 \\ (j,\alpha)\in S}}^{i-1}\sum_{\beta=1}^{\mathcal{L}}\sum_{\substack{k=1 \\ (k,\beta)\in S}}^{N}\sum_{\substack{r=1 \\ (r,\beta)\in S}}^{k-1}C_{ij\alpha}^{\text{con}}C_{kr\beta}^{\text{con}}\right]- \notag\\
&\sum_{\alpha=1}^{\mathcal{L}}\sum_{\substack{i=1 \\ (i,\alpha)\in S}}^{N}\sum_{\substack{j=1 \\ (j,\alpha)\in S}}^{i-1}\sum_{\beta=1}^{\mathcal{L}}\sum_{\substack{k=1 \\ (k,\beta)\in S}}^{N}\sum_{\substack{r=1 \\ (r,\beta)\in S}}^{k-1}C_{ij\alpha}C_{kr\beta} \notag\\
=&\sum_{\alpha=1}^{\mathcal{L}}\sum_{\substack{i=1 \\ (i,\alpha)\in S}}^{N}\sum_{\substack{j=1 \\ (j,\alpha)\in S}}^{i-1}\sum_{\beta=1}^{\mathcal{L}}\sum_{\substack{k=1 \\ (k,\beta)\in S}}^{N}\sum_{\substack{r=1 \\ (r,\beta)\in S}}^{k-1}\left(\text{E}[C_{ij\alpha}^{\text{con}}C_{kr\beta}^{\text{con}}]-C_{ij\alpha}C_{kr\beta}\right) \notag\\
{}=&\sum_{\alpha=1}^{\mathcal{L}}\sum_{\substack{i=1 \\ (i,\alpha)\in S}}^{N}\sum_{\substack{j=1 \\ (j,\alpha)\in S}}^{i-1}\sum_{\beta=1}^{\mathcal{L}}\sum_{\substack{k=1 \\ (k,\beta)\in S}}^{N}\sum_{\substack{r=1 \\ (r,\beta)\in S}}^{k-1}\left\{ \text{E}\left[\left(\frac{1}{L}\sum_{l=1}^{L}x_{il\alpha}x_{jl\alpha}\right)\left(\frac{1}{L}\sum_{l=1}^{L}x_{kl\beta}x_{rl\beta}\right)\right]- C_{ij\alpha}C_{kr\beta}\right\} \notag\\
=&\sum_{\alpha=1}^{\mathcal{L}}\sum_{\substack{i=1 \\ (i,\alpha)\in S}}^{N}\sum_{\substack{j=1 \\ (j,\alpha)\in S}}^{i-1}\sum_{\beta=1}^{\mathcal{L}}\sum_{\substack{k=1 \\ (k,\beta)\in S}}^{N}\sum_{\substack{r=1 \\ (r,\beta)\in S}}^{k-1}\left(\frac{1}{L^2}\text{E}\left[\sum_{l=1}^{L}\sum_{l^{\prime}=1}^{L}x_{il\alpha}x_{jl\alpha}x_{kl^{\prime}\beta}x_{rl^{\prime}\beta}\right]-C_{ij\alpha}C_{kr\beta}\right),\tag{S23}
\end{align}
where $L$ is the number of samples we draw from the $N$-variate multivariate normal distribution.
Now, using the fact that different samples are independent, we obtain
\begin{align} \label{eq:Vartotweight2}
& \frac{1}{L^2}\text{E}\left[\sum_{l=1}^{L}\sum_{l^{\prime}=1}^{L}x_{il\alpha}x_{jl\alpha}x_{kl^{\prime}\beta}x_{rl^{\prime}\beta}\right]-C_{ij\alpha}C_{kr\beta}\notag\\
=&\frac{1}{L^2}\sum_{l=1}^{L}\sum_{\substack{l^{\prime}=1 \\ l^{\prime}\neq l}}^{L}\text{E}[x_{il\alpha}x_{jl\alpha}x_{kl^{\prime}\beta}x_{rl^{\prime}\beta}]
+\frac{1}{L^2}\sum_{l=1}^{L}\text{E}[x_{il\alpha}x_{jl\alpha}x_{kl\beta}x_{rl\beta}]-C_{ij\alpha}C_{kr\beta} \notag\\
=&\frac{1}{L^2}\sum_{l=1}^{L}\text{E}[x_{il\alpha}x_{jl\alpha}]\cdot\sum_{\substack{l^{\prime}=1 \\ l^{\prime}\neq l}}^{L}\text{E}[x_{kl^{\prime}\beta}x_{rl^{\prime}\beta}]
+\frac{1}{L^2}\sum_{l=1}^{L}\text{E}[x_{il\alpha}x_{jl\alpha}x_{kl\beta}x_{rl\beta}]-C_{ij\alpha}C_{kr\beta} \notag\\
=&\frac{1}{L^2}\cdot LC_{ij\alpha}\cdot (L-1)C_{kr\beta}
+\frac{1}{L^2}\sum_{l=1}^{L}\text{E}[x_{il\alpha}x_{jl\alpha}x_{kl\beta}x_{rl\beta}]-C_{ij\alpha}C_{kr\beta} \notag\\
=&\frac{(L-1)}{L}C_{ij\alpha}C_{kr\beta}
+\frac{1}{L^2}\sum_{l=1}^{L}\text{E}[x_{il\alpha}x_{jl\alpha}x_{kl\beta}x_{rl\beta}]-C_{ij\alpha}C_{kr\beta} \notag\\
=&-\frac{1}{L}C_{ij\alpha}C_{kr\beta}+\frac{1}{L^2}\sum_{l=1}^{L}\text{E}[x_{il\alpha}x_{jl\alpha}x_{kl\beta}x_{rl\beta}].\tag{S24}
\end{align}
By substituting Eq.~\eqref{eq:Vartotweight2} into Eq.~\eqref{eq:Vartotweight1}, using the fact that the covariance matrices $C_{\alpha}$ and $C_{\beta}$ are independent when $\beta\neq\alpha$, and using Isserlis' Theorem \cite{isserlis1918formula}, we obtain 
\begin{align} \label{eq:Vartotweight3}
\text{Var}\left[\sum_{\alpha=1}^{\mathcal{L}}\sum_{\substack{i=1 \\ (i,\alpha)\in S}}^{N}\sum_{\substack{j=1 \\ (j,\alpha)\in S}}^{i-1}C_{ij\alpha}^{\text{con}}\right]=&\sum_{\alpha=1}^{\mathcal{L}}\sum_{\substack{i=1 \\ (i,\alpha)\in S}}^{N}\sum_{\substack{j=1 \\ (j,\alpha)\in S}}^{i-1}\sum_{\beta=1}^{\mathcal{L}}\sum_{\substack{k=1 \\ (k,\beta)\in S}}^{N}\sum_{\substack{r=1 \\ (r,\beta)\in S}}^{k-1}\left(-\frac{1}{L}C_{ij\alpha}C_{kr\beta}+\frac{1}{L^2}\sum_{l=1}^{L}\text{E}[x_{il\alpha}x_{jl\alpha}x_{kl\beta}x_{rl\beta}]\right) \notag\\
=&\sum_{\alpha=1}^{\mathcal{L}}\sum_{\substack{i=1 \\ (i,\alpha)\in S}}^{N}\sum_{\substack{j=1 \\ (j,\alpha)\in S}}^{i-1}\sum_{\substack{\beta=1 \\ \beta\neq\alpha}}^{\mathcal{L}}\sum_{\substack{k=1 \\ (k,\beta)\in S}}^{N}\sum_{\substack{r=1 \\ (r,\beta)\in S}}^{k-1}\left(-\frac{1}{L}C_{ij\alpha}C_{kr\beta}+\frac{1}{L^2}\sum_{l=1}^{L}\text{E}[x_{il\alpha}x_{jl\alpha}x_{kl\beta}x_{rl\beta}]\right) \notag\\
&+\sum_{\alpha=1}^{\mathcal{L}}\sum_{\substack{i=1 \\ (i,\alpha)\in S}}^{N}\sum_{\substack{j=1 \\ (j,\alpha)\in S}}^{i-1}\sum_{\substack{k=1 \\ (k,\alpha)\in S}}^{N}\sum_{\substack{r=1 \\ (r,\alpha)\in S}}^{k-1}\left(-\frac{1}{L}C_{ij\alpha}C_{kr\alpha}+\frac{1}{L^2}\sum_{l=1}^{L}\text{E}[x_{il\alpha}x_{jl\alpha}x_{kl\alpha}x_{rl\alpha}]\right) \notag\\
=&-\frac{1}{L}\left(\sum_{\alpha=1}^{\mathcal{L}}\sum_{\substack{i=1 \\ (i,\alpha)\in S}}^{N}\sum_{\substack{j=1 \\ (j,\alpha)\in S}}^{i-1}C_{ij\alpha}\right)\left(\sum_{\substack{\beta=1 \\ \beta\neq\alpha}}^{\mathcal{L}}\sum_{\substack{k=1 \\ (k,\beta)\in S}}^{N}\sum_{\substack{r=1 \\ (r,\beta)\in S}}^{k-1}C_{kr\beta}\right) \notag\\
&+\frac{1}{L^2}\sum_{l=1}^{L}\left(\sum_{\alpha=1}^{\mathcal{L}}\sum_{\substack{i=1 \\ (i,\alpha)\in S}}^{N}\sum_{\substack{j=1 \\ (j,\alpha)\in S}}^{i-1}\text{E}[x_{il\alpha}x_{jl\alpha}]\right)\left(\sum_{\substack{\beta=1 \\ \beta\neq\alpha}}^{\mathcal{L}}\sum_{\substack{k=1 \\ (k,\beta)\in S}}^{N}\sum_{\substack{r=1 \\ (r,\beta)\in S}}^{k-1}\text{E}[x_{kl\beta}x_{rl\beta}]\right) \notag\\
&-\frac{1}{L}\sum_{\alpha=1}^{\mathcal{L}}\left(\sum_{\substack{i=1 \\ (i,\alpha)\in S}}^{N}\sum_{\substack{j=1 \\ (j,\alpha)\in S}}^{i-1}C_{ij\alpha}\right)\left(\sum_{\substack{k=1 \\ (k,\alpha)\in S}}^{N}\sum_{\substack{r=1 \\ (r,\alpha)\in S}}^{k-1}C_{kr\alpha}\right) \notag\\
&+\frac{1}{L^2}\sum_{l=1}^{L}\left\{ \sum_{\alpha=1}^{\mathcal{L}}\sum_{\substack{i=1 \\ (i,\alpha)\in S}}^{N}\sum_{\substack{j=1 \\ (j,\alpha)\in S}}^{i-1}\sum_{\substack{k=1 \\ (k,\alpha)\in S}}^{N}\sum_{\substack{r=1 \\ (r,\alpha)\in S}}^{k-1}(\text{E}[x_{il\alpha}x_{jl\alpha}]\cdot\text{E}[x_{kl\alpha}x_{rl\alpha}] \right.\notag\\
&\left.+\text{E}[x_{il\alpha}x_{kl\alpha}]\cdot\text{E}[x_{jl\alpha}x_{rl\alpha}]+\text{E}[x_{il\alpha}x_{rl\alpha}]\cdot\text{E}[x_{jl\alpha}x_{kl\alpha}])\right\} \notag\\
=&-\frac{1}{L}\left(\sum_{\alpha=1}^{\mathcal{L}}\sum_{\substack{i=1 \\ (i,\alpha)\in S}}^{N}\sum_{\substack{j=1 \\ (j,\alpha)\in S}}^{i-1}C_{ij\alpha}\right)\left(\sum_{\substack{\beta=1 \\ \beta\neq\alpha}}^{\mathcal{L}}\sum_{\substack{k=1 \\ (k,\beta)\in S}}^{N}\sum_{\substack{r=1 \\ (r,\beta)\in S}}^{k-1}C_{kr\beta}\right) \notag\\
&+\frac{1}{L^2}\cdot L\left(\sum_{\alpha=1}^{\mathcal{L}}\sum_{\substack{i=1 \\ (i,\alpha)\in S}}^{N}\sum_{\substack{j=1 \\ (j,\alpha)\in S}}^{i-1}C_{ij\alpha}\right)\left(\sum_{\substack{\beta=1 \\ \beta\neq\alpha}}^{\mathcal{L}}\sum_{\substack{k=1 \\ (k,\beta)\in S}}^{N}\sum_{\substack{r=1 \\ (r,\beta)\in S}}^{k-1}C_{kr\beta}\right) \notag\\
&-\frac{1}{L}\sum_{\alpha=1}^{\mathcal{L}}\left(\sum_{\substack{i=1 \\ (i,\alpha)\in S}}^{N}\sum_{\substack{j=1 \\ (j,\alpha)\in S}}^{i-1}C_{ij\alpha}\right)\left(\sum_{\substack{k=1 \\ (k,\alpha)\in S}}^{N}\sum_{\substack{r=1 \\ (r,\alpha)\in S}}^{k-1}C_{kr\alpha}\right) \notag\\
&+\frac{1}{L^2}\cdot L\left[\sum_{\alpha=1}^{\mathcal{L}}\sum_{\substack{i=1 \\ (i,\alpha)\in S}}^{N}\sum_{\substack{j=1 \\ (j,\alpha)\in S}}^{i-1}\sum_{\substack{k=1 \\ (k,\alpha)\in S}}^{N}\sum_{\substack{r=1 \\ (r,\alpha)\in S}}^{k-1}(C_{ij\alpha}C_{kr\alpha}+C_{ik\alpha}C_{jr\alpha}+C_{ir\alpha}C_{jk\alpha})\right].\tag{S25}
\end{align}
By combining the first and third terms in Eq.~\eqref{eq:Vartotweight3}, we obtain
\begin{align} \label{eq:Vartotweight5}
\text{Var}\left[\sum_{\alpha=1}^{\mathcal{L}}\sum_{\substack{i=1 \\ (i,\alpha)\in S}}^{N}\sum_{\substack{j=1 \\ (j,\alpha)\in S}}^{i-1}C_{ij\alpha}^{\text{con}}\right] =& -\frac{1}{L}\left(\sum_{\alpha=1}^{\mathcal{L}}\sum_{\substack{i=1 \\ (i,\alpha)\in S}}^{N}\sum_{\substack{j=1 \\ (j,\alpha)\in S}}^{i-1}C_{ij\alpha}\right)\left(\sum_{\beta=1}^{\mathcal{L}}\sum_{\substack{k=1 \\ (k,\beta)\in S}}^{N}\sum_{\substack{r=1 \\ (r,\beta)\in S}}^{k-1}C_{kr\beta}\right) \notag\\
&+\frac{1}{L}\left(\sum_{\alpha=1}^{\mathcal{L}}\sum_{\substack{i=1 \\ (i,\alpha)\in S}}^{N}\sum_{\substack{j=1 \\ (j,\alpha)\in S}}^{i-1}C_{ij\alpha}\right)\left(\sum_{\substack{\beta=1 \\ \beta\neq\alpha}}^{\mathcal{L}}\sum_{\substack{k=1 \\ (k,\beta)\in S}}^{N}\sum_{\substack{r=1 \\ (r,\beta)\in S}}^{k-1}C_{kr\beta}\right) \notag\\
&+\frac{1}{L}\left[\sum_{\alpha=1}^{\mathcal{L}}\sum_{\substack{i=1 \\ (i,\alpha)\in S}}^{N}\sum_{\substack{j=1 \\ (j,\alpha)\in S}}^{i-1}\sum_{\substack{k=1 \\ (k,\alpha)\in S}}^{N}\sum_{\substack{r=1 \\ (r,\alpha)\in S}}^{k-1}(C_{ij\alpha}C_{kr\alpha}+C_{ik\alpha}C_{jr\alpha}+C_{ir\alpha}C_{jk\alpha})\right] \notag\\
=& -\frac{1}{L}\left(\sum_{\alpha=1}^{\mathcal{L}}\sum_{\substack{i=1 \\ (i,\alpha)\in S}}^{N}\sum_{\substack{j=1 \\ (j,\alpha)\in S}}^{i-1} C_{ij\alpha}\right)^{2} \notag\\
&+\frac{1}{L}\left(\sum_{\alpha=1}^{\mathcal{L}}\sum_{\substack{i=1 \\ (i,\alpha)\in S}}^{N}\sum_{\substack{j=1 \\ (j,\alpha)\in S}}^{i-1}C_{ij\alpha}\right)\left(\sum_{\substack{\beta=1 \\ \beta\neq\alpha}}^{\mathcal{L}}\sum_{\substack{k=1 \\ (k,\beta)\in S}}^{N}\sum_{\substack{r=1 \\ (r,\beta)\in S}}^{k-1}C_{kr\beta}\right) \notag\\
&+\frac{1}{L}\left(\sum_{\alpha=1}^{\mathcal{L}}\sum_{\substack{i=1 \\ (i,\alpha)\in S}}^{N}\sum_{\substack{j=1 \\ (j,\alpha)\in S}}^{i-1}\sum_{\substack{k=1 \\ (k,\alpha)\in S}}^{N}\sum_{\substack{r=1 \\ (r,\alpha)\in S}}^{k-1}C_{ij\alpha}C_{kr\alpha}\right) \notag\\
&+\frac{1}{L}\left[\sum_{\alpha=1}^{\mathcal{L}}\sum_{\substack{i=1 \\ (i,\alpha)\in S}}^{N}\sum_{\substack{j=1 \\ (j,\alpha)\in S}}^{i-1}\sum_{\substack{k=1 \\ (k,\alpha)\in S}}^{N}\sum_{\substack{r=1 \\ (r,\alpha)\in S}}^{k-1}(C_{ik\alpha}C_{jr\alpha}+C_{ir\alpha}C_{jk\alpha})\right] \notag\\
=& -\frac{1}{L}\left(\sum_{\alpha=1}^{\mathcal{L}}\sum_{\substack{i=1 \\ (i,\alpha)\in S}}^{N}\sum_{\substack{j=1 \\ (j,\alpha)\in S}}^{i-1} C_{ij\alpha}\right)^{2}+\frac{1}{L}\left(\sum_{\alpha=1}^{\mathcal{L}}\sum_{\substack{i=1 \\ (i,\alpha)\in S}}^{N}\sum_{\substack{j=1 \\ (j,\alpha)\in S}}^{i-1} C_{ij\alpha}\right)^{2} \notag\\
&+\frac{1}{L}\left[\sum_{\alpha=1}^{\mathcal{L}}\sum_{\substack{i=1 \\ (i,\alpha)\in S}}^{N}\sum_{\substack{j=1 \\ (j,\alpha)\in S}}^{i-1}\sum_{\substack{k=1 \\ (k,\alpha)\in S}}^{N}\sum_{\substack{r=1 \\ (r,\alpha)\in S}}^{k-1}(C_{ik\alpha}C_{jr\alpha}+C_{ir\alpha}C_{jk\alpha})\right] \notag\\
=& \frac{1}{L}\left[\sum_{\alpha=1}^{\mathcal{L}}\sum_{\substack{i=1 \\ (i,\alpha)\in S}}^{N}\sum_{\substack{j=1 \\ (j,\alpha)\in S}}^{i-1}\sum_{\substack{k=1 \\ (k,\alpha)\in S}}^{N}\sum_{\substack{r=1 \\ (r,\alpha)\in S}}^{k-1}(C_{ik\alpha}C_{jr\alpha}+C_{ir\alpha}C_{jk\alpha})\right].\tag{S26}
\end{align}

\subsection*{Text D. Graphical lasso} \label{graphicallasso}

For multivariate Gaussian distributions, a zero in the precision matrix (i.e., inverse covariance matrix) is equivalent to conditional independence of two variables, which one can relate to the absence of an edge in the network. Therefore, it is natural to use the non-zero entries of the estimated precision matrix to determine the edges \cite{hsieh2014quic}. However, when the number of variables is larger than the number of samples, the empirical covariance matrix is not full rank. In this case, the empirical covariance matrix is singular, meaning that its condition number is infinite, so estimating the precision matrix becomes difficult \cite{friedman2008sparse}. The graphical lasso addresses this problem by regularizing the maximum likelihood estimator with a lasso penalty enforcing sparsity \cite{meinshausen2006high, yuan2007model, friedman2008sparse}.

Let $\Vec{y}$ be a $p$-variate Gaussian random column vector, with distribution $\mathcal{N}(\mu,C)$, where $\mu$ is the $p$-dimensional mean vector and $C$ is the $p\times p$ covariance matrix. Given $n$ independently drawn samples $\{\Vec{y_1},\ldots,\Vec{y_n}\}$ of this random vector, the sample covariance matrix can be written as
\begin{equation} \label{eq:sampcov}
\hat{C}=\frac{1}{n-1}\sum_{k=1}^{n}(\Vec{y_k}-\hat{\mu})(\Vec{y_k}-\hat{\mu})^{\top},\tag{S27}
\end{equation}
where $\hat{\mu}=\frac{1}{n}\sum_{k=1}^{n}\Vec{y_{k}}$, and ${}^{\top}$ represents the transposition.
Let the inverse covariance matrix be denoted as $C^{-1}=\Theta$. We consider a generalized $\ell_{1}$ regularization given by $\lambda\sum_{i=1}^{p}\sum_{j=1}^{i-1}|\Theta_{ij}|$, where $\lambda$ is the penalizing parameter. Then, the problem is to maximize the lasso regularized log-likelihood to obtain the graphical lasso estimator, i.e.,
\begin{equation} \label{eq:lasso1}
\Theta^{*}=\text{arg}\min_{\Theta\succ 0}\left\{-\log\det \Theta+\text{tr}(\hat{C}\Theta)+\lambda\sum_{i=1}^p \sum_{j=1}^{i-1}|\Theta_{ij}|\right\},\tag{S28}
\end{equation}
where $\Theta \succ 0$ signifies that $\Theta$ is a positive definite matrix \cite{friedman2008sparse}.

Using the TPM data for each of the four tissues separately, we first calculated the $203\times 203$ empirical covariance matrix. Then, we applied the GraphicalLassoCV function from the Python package scikit-learn version 1.0.2 \cite{scikit-learn, scikit-learn-code} to estimate a precision matrix, which is a sparsified co-expression network, for each tissue. This graphical lasso algorithm incorporates a cross-validated choice of the $\ell_{1}$ penalty. To simplify analysis, we regard the generated networks as unsigned and unweighted network. The results for the unsigned weighted networks are similar to those for the unsigned unweighted networks, as we will show.

\subsection*{Text E. Z scores for the average distance between pairs of genes on each chromosome separately in each community} \label{SI3}

We show in Table~\ref{table:comm_by_chrom_zscores1} the Z scores for the average distance between pairs of genes on each chromosome and each community with $\gamma=1$. We show the corresponding results with $\gamma=3$ in Table~\ref{table:comm_by_chrom_zscores3}. We only calculated the Z scores for the chromosome-community pairs with at least 3 genes. In these tables, N/A implies that either there are less than 3 genes, or the standard deviation of the average distance is equal to 0 because every random selection of genes is the same gene set. The latter event occurs when all the genes on that chromosome are associated with the same community. 

\begin{table}[t]
\renewcommand\thetable{B}
\begin{center}
\caption{Z scores for the average distance between pairs of genes on each chromosome and each significant community detected with $\gamma=1$. Comm.\,denotes community, and Chr denotes chromosome.}
\begin{tabular}[b]{|c | c c c c c|}\hline
\multicolumn{6}{|c|}{$\gamma=1$} \\\hline
& \multicolumn{5}{|c|}{Comm.} \\\hline
Chr & 1    & 2    & 3   & 4   & 5   \\\hline
1          &      $-2.093$&     $-1.526$ &     $-0.122$&     $-1.466$&   $1.272$  \\
2          &      N/A&      $-0.767$&     $0.061$&     $2.129$&  $-0.725$   \\
3          &      N/A&      N/A&     $0.113$&     $1.101$&   N/A  \\
4          &      N/A&     N/A &     N/A&     $-0.110$&  $-2.573$   \\
5          &      N/A&      N/A&     N/A&     N/A& N/A    \\
6          &      $0.606$&      N/A&     N/A&     $0.564$&   $-0.065$  \\
7          &      N/A&      $-1.938$&     $-1.842$&     $0.968$&   $-0.302$  \\
8          &      N/A&      N/A&     N/A&     N/A&  N/A   \\
9          &      N/A&      N/A&     N/A&     $-0.722$&   N/A  \\
10         &      N/A&      $-2.336$&     $0.413$&     N/A&   N/A  \\
11         &      N/A&      N/A&     $-2.156$&     $0.077$&   $0.454$  \\
12         &      N/A&      N/A&     $-1.376$&     $-0.905$&   $-0.260$  \\
13         &      N/A&      N/A&     N/A&     N/A&   N/A  \\
14         &      N/A&      N/A&     $0.438$&     N/A&  $0.180$   \\
15         &      N/A&      N/A&     N/A&     N/A&   N/A  \\
16         &      $-0.091$&      $0.737$&     $0.100$&     N/A&    N/A \\
17         &      $1.535$&      N/A&     $-2.160$&     $-0.579$&  $-0.787$   \\
18         &      N/A&      N/A&     N/A&     N/A&   N/A  \\
19         &      $0.235$&      N/A&     $-0.521$&     $1.592$&  $1.894$   \\
20         &      N/A&      N/A&     N/A&     N/A&  $0.791$   \\
21         &      N/A&      N/A&     N/A&     N/A&   N/A  \\
22         &      N/A&      N/A&     N/A&     N/A&  $-1.780$   \\
X          &      N/A&      N/A&     N/A&     N/A&  N/A   \\
Y          &      N/A&      N/A&     N/A&     N/A&  N/A   \\
M          &      N/A&      N/A&     N/A&     N/A&  $-2.061$  \\\hline
 \end{tabular}
 \label{table:comm_by_chrom_zscores1}
 \end{center}
 \end{table}
 
\begin{table}[t]
\renewcommand\thetable{C}
\begin{center}
\caption{Z scores for the average distance between pairs of genes on each chromosome and each significant community detected with $\gamma=3$. Comm.\,denotes community, and Chr denotes chromosome.}
\begin{tabular}[b]{|c | c c c c c c c|}\hline
\multicolumn{8}{|c|}{$\gamma=3$} \\\hline
& \multicolumn{7}{|c|}{Comm.} \\\hline
Chr & 1  & 2  & 3  & 4  & 5 & 6 & 7 \\\hline
1          &    $-2.808$&    $0.116$&    $-0.447$&    N/A&   $-0.482$&   N/A&  N/A \\
2          &    $-0.671$&    N/A&    N/A&    N/A&   $2.094$&   $-2.220$&  N/A \\
3          &    N/A&    N/A&    N/A&    N/A&   N/A&   N/A&  N/A \\
4          &    N/A&    N/A&    N/A&    N/A&   $-1.993$&   N/A& N/A  \\
5          &    N/A&    N/A&    N/A&    N/A&   N/A&   N/A&  N/A \\
6          &    N/A&   $0.573$&    N/A&    N/A&   N/A&   N/A&  N/A \\
7          &    $-1.643$&    N/A&    N/A&    N/A&   $-2.616$&   N/A&  N/A \\
8          &    N/A&    N/A&    N/A&    N/A&   N/A&   N/A&  N/A \\
9          &    N/A&    N/A&    N/A&    N/A&   N/A&   N/A&  N/A \\
10         &    N/A&    N/A&    N/A&    N/A&   $0.293$&   N/A&  N/A \\
11         &    N/A&    N/A&    N/A&    N/A&   $-2.902$&   N/A&  N/A \\
12         &    N/A&    $1.729$&    N/A&    N/A&   $-0.577$&   N/A&  N/A \\
13         &    N/A&    N/A&    N/A&    N/A&   N/A&   N/A&  N/A \\
14         &    N/A&    N/A&    N/A&    N/A&   N/A&   $-1.756$&  N/A \\
15         &    N/A&    N/A&    N/A&    N/A&   N/A&   N/A&  N/A \\
16         &    $-0.416$&    $0.920$&    N/A&    N/A&   $1.016$&   N/A& N/A  \\
17         &    N/A&    $1.639$&    N/A&    N/A&   $-2.175$&   N/A&  N/A \\
18         &    N/A&    N/A&    N/A&    N/A&   N/A&   N/A&  N/A \\
19         &    N/A&    $-1.142$&    N/A&    N/A&   $-0.179$&   N/A&  N/A \\
20         &    N/A&    N/A&    N/A&    N/A&   N/A&   N/A&  N/A \\
21         &    N/A&    N/A&    N/A&    N/A&   N/A&   N/A&  N/A \\
22         &    N/A&    N/A&    N/A&    N/A&   N/A&   N/A&  N/A \\
X          &    N/A&    N/A&    N/A&    N/A&   N/A&   N/A&  N/A \\
Y          &    N/A&    N/A&    N/A&    N/A&   N/A&   N/A&  N/A \\
M          &    N/A&    N/A&    N/A&    N/A&   N/A&   N/A&  N/A \\\hline
 \end{tabular}
 \label{table:comm_by_chrom_zscores3}
 \end{center}
 \end{table}
 
 \newpage

 \subsection*{Text F. Results of the gene set enrichment analysis} \label{SI4}

We show the two most significant GO:BP and HP results from g:Profiler for the top 50 highly expressed genes out of the $203$ genes in the network in each tissue in Table~\ref{table:gProf_highlyexp}. See \cite{gpropanc} for the entire output from g:Profiler, i.e., the list of all significant GO:BP and HP results, for the top 50 genes in pancreas; see \cite{gprosal} for the top 50 genes in salivary gland; see \cite{gpromamm} for the top 50 genes in mammary gland; and see \cite{gproskin} for the top 50 genes in skin.

We show the two most significant GO:BP and HP results from g:Profiler for each community in the partition of the multilayer correlation matrix with $\gamma=3$ in Table~\ref{table:gProf_comms}. See \cite{gpro1} for the entire output from g:Profiler, i.e., the list of all significant GO:BP and HP results, for the genes in community 1; see \cite{gpro2} for community 2; see \cite{gpro3} for community 3; see \cite{gpro4} for community 4; see \cite{gpro5} for community 5; see \cite{gpro6} for community 6; and see \cite{gpro7} for community 7.

We show the two most significant GO:BP and HP results from g:Profiler for the top 50 highly connected genes (i.e., top 50 hub genes) out of the $203$ genes in the network in each tissue in Table~\ref{table:gProf_hubs}. See \cite{gpropanchubs} for the entire output from g:Profiler, i.e., the list of all significant GO:BP and HP results, for the top 50 genes in pancreas; see \cite{gprosalhubs} for the top 50 genes in salivary gland; see \cite{gpromammhubs} for the top 50 genes in mammary gland; and see \cite{gproskinhubs} for the top 50 genes in skin.

\begin{table}[t]
 \renewcommand\thetable{D}
\begin{center}
 \caption{Results of the gene set enrichment analysis for the top 50 highly expressed genes out of the $203$ genes in the network in each tissue.}
\begin{tabular}[b]{|c c c c c|}\hline
Tissue & \thead{Top significant \\ terms from GO:BP} & $p$ value & \thead{Top significant \\ terms from HP} & $p$ value\\ \hline
  \multirow{2}{*}{pancreas} & oxidative phosphorylation & $6.39\cdot 10^{-15}$ & recurrent pancreatitis & $1.44\cdot 10^{-22}$\\
  & aerobic electron transport chain & $4.69\cdot 10^{-14}$ & mitochondrial inheritance & $1.79\cdot 10^{-20}$\\\hline
  \multirow{2}{*}{salivary gland} & oxidative phosphorylation & $7.02\cdot 10^{-16}$ & mitochondrial inheritance & $6.33\cdot 10^{-21}$\\
  & aerobic electron transport chain & $7.37\cdot 10^{-15}$ & centrocecal scotoma & $3.23\cdot 10^{-20}$\\\hline
  \multirow{3}{*}{mammary gland} & oxidative phosphorylation & $6.35\cdot 10^{-15}$ & mitochondrial inheritance & $9.67\cdot 10^{-20}$ \\ 
  & \thead{mitochondrial ATP synthesis \\ coupled electron transport} & $4.77\cdot 10^{-14}$ & centrocecal scotoma & $3.63\cdot 10^{-19}$\\\hline
  \multirow{2}{*}{skin} & oxidative phosphorylation & $7.28\cdot 10^{-15}$ & mitochondrial inheritance & $5.06\cdot 10^{-19}$\\
  & aerobic electron transport chain & $5.35\cdot 10^{-14}$ & centrocecal scotoma & $1.59\cdot 10^{-18}$\\\hline
 \end{tabular}
 \label{table:gProf_highlyexp}
 \end{center}
 \end{table}

\begin{table}[t]
\renewcommand\thetable{E}
\begin{center}
 \caption{Results of the gene set enrichment analysis for the communities of the multilayer correlation matrix with $\gamma=3$. Comm.\,denotes community.}
\begin{tabular}[b]{|c c c c c|}\hline
Comm. & \thead{Top significant \\ terms from GO:BP} & $p$ value & \thead{Top significant \\ terms from HP} & $p$ value\\ \hline
  \multirow{2}{*}{1} & oxidative phosphorylation & $5.90\cdot 10^{-11}$ & recurrent pancreatitis & $1.78\cdot 10^{-19}$\\
  & aerobic electron transport chain & $1.05\cdot 10^{-10}$ & mitochondrial inheritance & $1.06\cdot 10^{-17}$\\\hline
  \multirow{2}{*}{2} & keratinocyte differentiation & $8.26\cdot 10^{-25}$ & palmoplantar keratoderma & $3.05\cdot 10^{-7}$\\
  & epidermis development & $1.90\cdot 10^{-24}$ & hyperkeratosis & $3.05\cdot 10^{-7}$\\\hline
  \multirow{2}{*}{3} & retina homeostasis & $2.52\cdot 10^{-10}$ & leber optic atophy & $6.65\cdot 10^{-15}$\\ 
  & oxidative phosphorylation & $2.23\cdot 10^{-9}$ & mitochondrial inheritance & $6.65\cdot 10^{-15}$\\\hline
  \multirow{2}{*}{4} & skin development & $3.23\cdot 10^{-13}$ & alopecia & $8.30\cdot 10^{-5}$\\
  & intermediate filament organization & $2.64\cdot 10^{-12}$ & nail dystrophy & $8.30\cdot 10^{-5}$\\\hline
  \multirow{2}{*}{5} & keratinization & $1.95\cdot 10^{-19}$ & palmoplantar blistering & $1.71\cdot 10^{-6}$\\
  & epidermis development & $1.95\cdot 10^{-19}$ & palmoplantar keratoderma & $2.02\cdot 10^{-5}$\\\hline
  \multirow{2}{*}{6} & positive regulation of respiratory burst & $5.36\cdot 10^{-8}$ & N/A & N/A\\
  & regulation of respiratory burst & $2.60\cdot 10^{-7}$ & N/A & N/A\\\hline
  \multirow{2}{*}{7} & adaptive thermogenesis & $1.73\cdot 10^{-2}$ & N/A & N/A\\
  & fatty acid biosynthesis process & $1.73\cdot 10^{-2}$ & N/A & N/A\\\hline
 \end{tabular}
 \label{table:gProf_comms}
 \end{center}
 \end{table}

\begin{table}[t]
 \renewcommand\thetable{F}
\begin{center}
 \caption{Results of the gene set enrichment analysis for the top 50 highly connected genes out of the $203$ genes in the single-layer network of each tissue.}
\begin{tabular}[b]{|c c c c c|}\hline
Tissue & \thead{Top significant \\ terms from GO:BP} & $p$ value & \thead{Top significant \\ terms from HP} & $p$ value\\ \hline
  \multirow{2}{*}{pancreas} & cytoplasmic translation & $1.68\cdot 10^{-10}$ & mutism & $1.77\cdot 10^{-3}$\\
  & sequestering of metal ion & $2.49\cdot 10^{-8}$ & arterial rupture & $1.93\cdot 10^{-3}$\\\hline
  \multirow{2}{*}{salivary gland} & digestion & $3.05\cdot 10^{-8}$ & pancreatic calcification & $5.28\cdot 10^{-9}$\\
  & cytoplasmic translation & $4.52\cdot 10^{-5}$ & pancreatic pseudocyst & $5.66\cdot 10^{-8}$\\\hline
  \multirow{2}{*}{mammary gland} & antibacterial humoral response & $3.23\cdot 10^{-10}$ & nail dystrophy & $9.67\cdot 10^{-5}$ \\ 
  & defense response to bacterium & $3.36\cdot 10^{-8}$ & palmoplantar blistering & $4.37\cdot 10^{-4}$\\\hline
  \multirow{2}{*}{skin} & digestion & $1.59\cdot 10^{-8}$ & pancreatic calcification & $4.68\cdot 10^{-6}$\\
  & proteolysis & $4.89\cdot 10^{-6}$ & recurrent pancreatitis & $4.82\cdot 10^{-5}$\\\hline
 \end{tabular}
 \label{table:gProf_hubs}
 \end{center}
 \end{table}
 
 \newpage

\subsection*{Text G. Tissue-specific hub genes versus gene communities} \label{SI7}

To compare the overlap of the top 50 highly connected genes in each tissue and the gene communities identified by our algorithm with $\gamma=3$, we calculate the Jaccard index for each pair of the set of the 50 most connected genes in one of the four tissues and one of the seven gene communities identified by our algorithm. 
The Jaccard index is equal to $1$ if the two sets perfectly overlap and $0$ if the two sets are disjoint. We show the $4\times 7 = 28$ Jaccard index values in Fig~\ref{fig:Jacc_inds}. The largest Jaccard index is $0.324$, revealing the lack of notable similarity between all of the $28$ pairs of gene sets. Therefore, we conclude that our multilayer community detection method uncovers sets of genes that are different from top hub genes in each layer.
 
 \renewcommand{\thefigure}{B}
\captionsetup{font={small,rm}} 
\captionsetup{labelfont=bf}
 \begin{figure}[]
  \begin{center}
    \includegraphics[width=\textwidth]{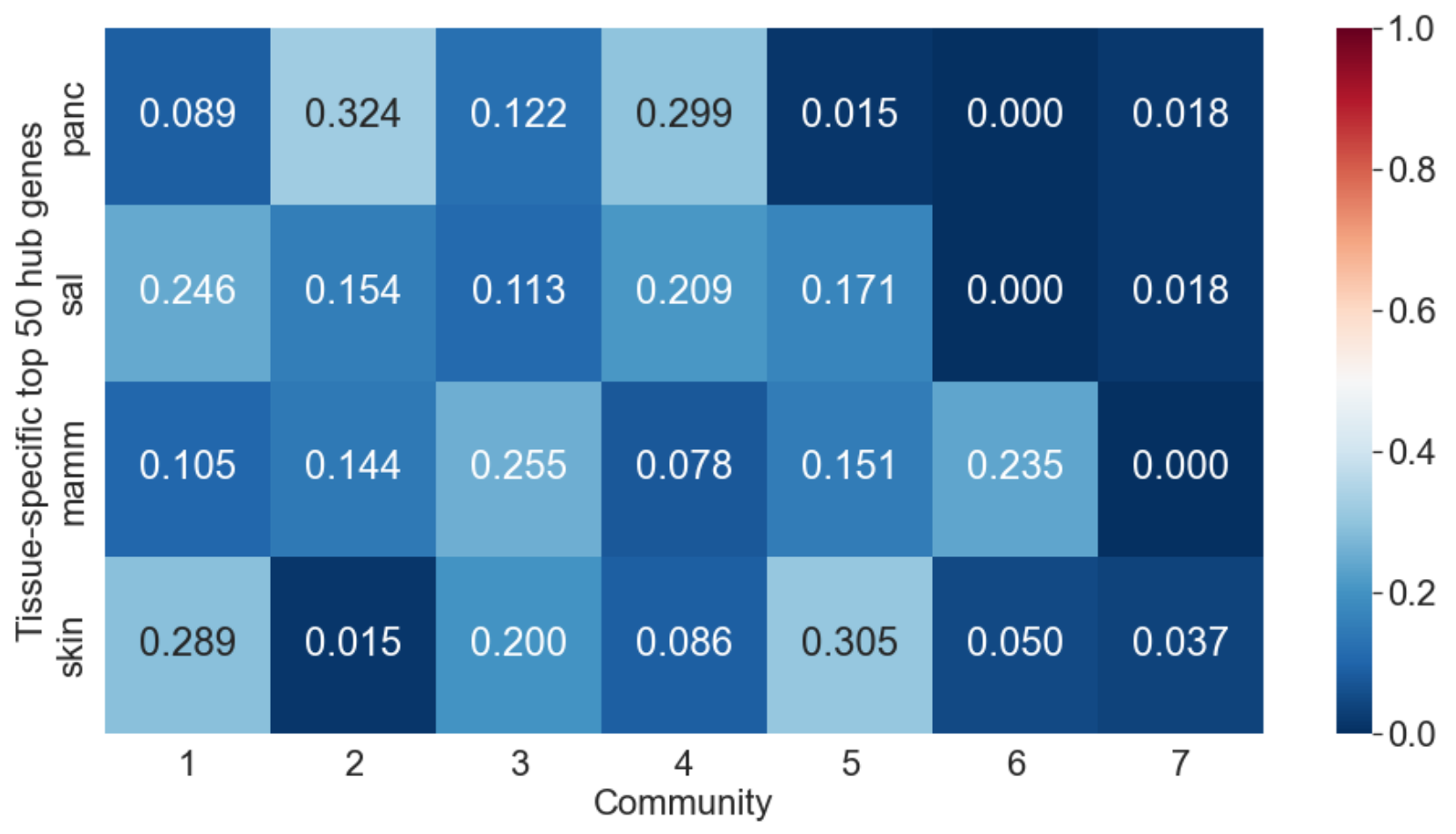}
  \end{center}
  \caption{Jaccard index between the set of tissue-specific hub genes and the set of genes in a community. Each row corresponds to the top 50 hub genes in each layer (i.e., tissue), where ``panc'' denotes pancreas, ``sal'' denotes salivary gland, ``mamm'' denotes mammary gland, and ``skin'' denotes skin (not sun exposed). Each column corresponds to a community identified with $\gamma=3$.}
  \label{fig:Jacc_inds}
\end{figure}

\clearpage

\renewcommand{\refname}{Supplementary References}

\end{document}